\documentclass{article}

    \PassOptionsToPackage{numbers, compress}{natbib}
 \usepackage[dblblindworkshop, final]{neurips_2026}
\workshoptitle{SocialAgent: Second Workshop on Large Language Models for Social Reasoning and Simulation}

\usepackage[utf8]{inputenc} 
\usepackage[T1]{fontenc}    
\usepackage{hyperref}       
\usepackage{url}            
\usepackage{booktabs}       
\usepackage{amsfonts}       
\usepackage{nicefrac}       
\usepackage{microtype}      
\usepackage{xcolor}         
\usepackage{graphicx}
\usepackage{amsmath}
\usepackage{float}
\usepackage{xurl}
\usepackage{enumitem}
\usepackage{url}
\usepackage{array}

\usepackage[most]{tcolorbox}
\usepackage{xcolor}

\definecolor{linegray}{RGB}{245,245,245}
\definecolor{redhl}{RGB}{255,220,220}
\definecolor{greenhl}{RGB}{220,245,220}

\tcbset{
  promptstyle/.style={
    colback=linegray,
    colframe=white,
    boxrule=0pt,
    sharp corners,
    left=6pt,
    right=6pt,
    top=4pt,
    bottom=4pt,
    fontupper=\ttfamily\scriptsize,
    before upper=\raggedright,
    breakable,
  }
}
\newtcolorbox{promptbox}[1][]{promptstyle,#1}

\newtcbox{\redhlbox}{
  on line,
  colback=redhl,
  colframe=redhl,
  arc=3pt,
  boxrule=0pt,
  boxsep=0.5pt,
  left=2pt,
  right=2pt,
  top=1pt,
  bottom=1pt
}

\newtcbox{\greenhlbox}{
  on line,
  colback=greenhl,
  colframe=greenhl,
  arc=3pt,
  boxrule=0pt,
  boxsep=0.5pt,
  left=2pt,
  right=2pt,
  top=1pt,
  bottom=1pt
}

\title{Stop Drawing Scientific Claims from LLM Social Simulations Without Robustness Audits}

\author{%
  Jinyi Ye\textsuperscript{1}\thanks{Equal contribution.} \quad
  Lei Cao\textsuperscript{2}\footnotemark[1] \quad
  Ding Chen\textsuperscript{3} \quad
  Emilio Ferrara\textsuperscript{1,2} \\
  \textsuperscript{1}Thomas Lord Department of Computer Science, University of Southern California \\
  \textsuperscript{2}Annenberg School for Communication and Journalism, University of Southern California \\
  \textsuperscript{3}Marshall School of Business, University of Southern California \\
  \texttt{\{jinyiy, caolei, blairche, emiliofe\}@usc.edu}
}

\begin{document}

\maketitle
\raggedbottom

\begin{abstract}
\textbf{The scientific claims drawn from LLM social simulations should be no stronger than the robustness audits that support them.} Generative agents bring new expressive power to agent-based modeling, enabling simulations of collective social processes like cooperation, polarization, and norm formation. Yet they also introduce complexity through additional architectural choices, such as agent specification, memory representation, interaction protocols, and environment design. Small perturbations that appear minor to researchers can cascade into macro-level outcomes through repeated interaction, creating a ``\textit{butterfly effect}.'' Consequently, scientific claims drawn from LLM social simulations may reflect implementation artifacts rather than the social mechanisms being modeled. We support this position with two case studies: a repeated Prisoner's Dilemma and a social media echo chamber simulation. Across multiple models, minor perturbations in persona format and game-instruction framing shift cooperation rates by up to $76$ percentage points, while network homophily and hub assignment produce significant and consistent shifts in polarization metrics. We also find that sensitivity is unevenly distributed across both architectural choices and model families: the same perturbation that produces the $76$ pp shift in one frontier model only shifts another by $1$ pp. Robustness is therefore a property that should be measured per claim and per model, not assumed. To address this validation gap, we introduce \textbf{TRAILS} (\textit{Taxonomy for Robustness Audits In LLM Simulations}), a robustness-audit taxonomy spanning three levels of simulation design: agent (micro-level), interaction (meso-level), and system (macro-level), together with a Social Simulation Audit Card for interpreting sensitivity and scoping scientific claims. We call for robustness to become a first-order validation requirement before LLM social simulations are used to explain mechanisms, evaluate interventions, or inform decisions. 
Code and data are available at \url{https://github.com/angelayejinyi/butterfly-effect-sim/tree/position}.
\end{abstract}


\section{Introduction}
\label{sec:intro}

\begin{figure}[t]
    \centering
    \includegraphics[width=0.95\textwidth]{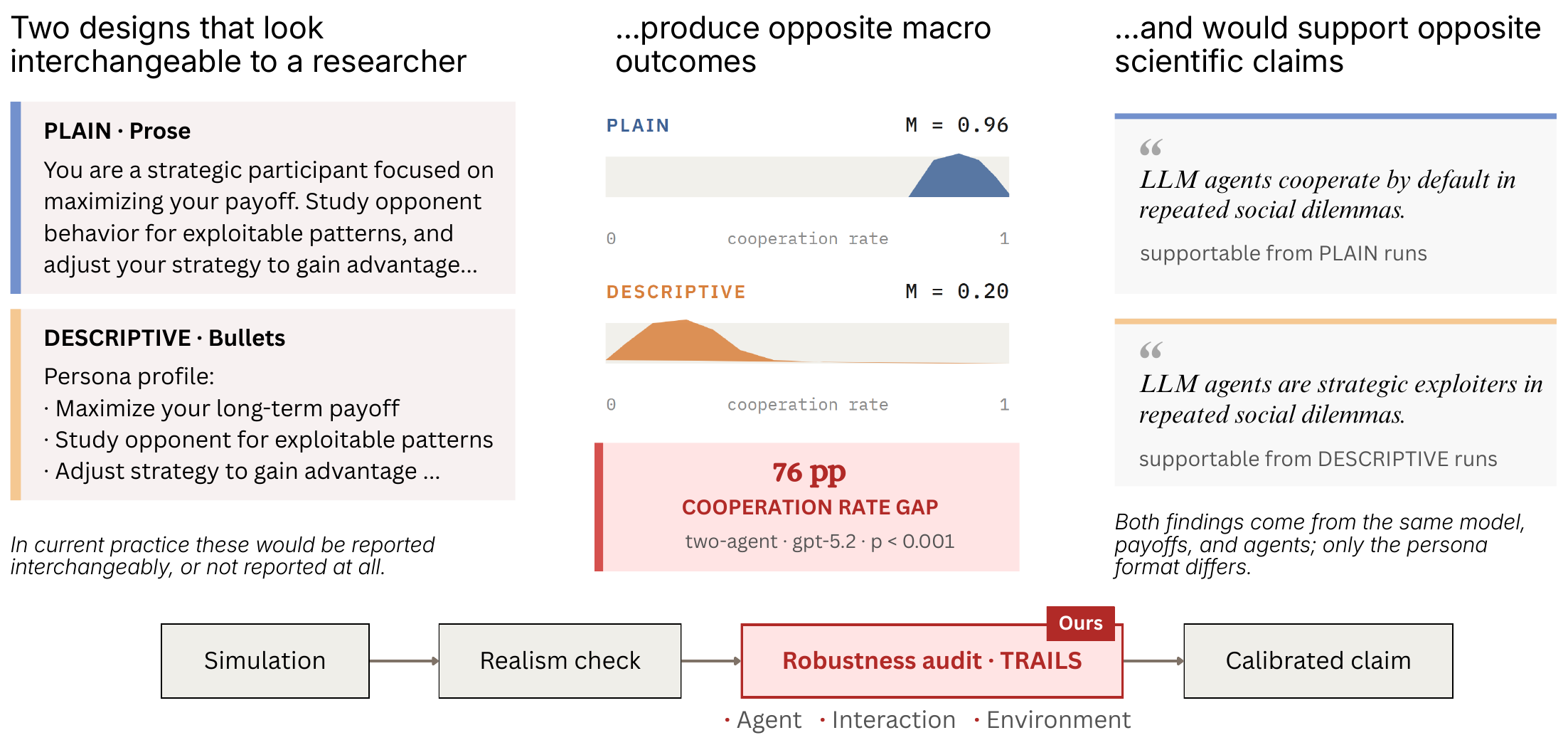}
    \caption{\textbf{The butterfly effect in LLM social simulations.}
    Two persona prompts that differ only in surface format while preserving the same content produce a 76-percentage-point gap in cooperation rate in a 10-round Prisoner's Dilemma (\texttt{gpt-5.2}, $N=30$ seeds per condition; two-sided Mann--Whitney $U$, $p<0.001$). We argue that LLM social simulations should not support claims stronger than their robustness audits can justify. We introduce \textbf{TRAILS}, a taxonomy for auditing simulations across three levels of design: agent (micro), interaction (meso), and system (macro). Section \ref{sec:case_study} demonstrates the effect across two case studies and four LLMs; Section \ref{sec:taxonomy} presents TRAILS.}
\label{fig:butterfly_effect_illustration}
\end{figure}

\textbf{LLM Social Simulations for Collective Human Behavior Modeling.}
Until recently, simulating collective human behavior in agent-based models required researchers to encode every social tendency by hand: how an agent decides to cooperate, who it talks to, what it remembers, and how its preferences change. Large language models have lifted this constraint. By generating agent behavior directly from natural language, they have opened a new design space for studying a wide range of social phenomena \citep{park2023generativeagents}. The field has moved quickly to occupy this space, and the purposes for which these simulations are deployed have escalated alongside it. Researchers now use LLM-based social simulations to \textit{explore and explain} social mechanisms such as cooperation \citep{akata2025repeatedgames}, opinion dynamics \citep{chuang2024simulating} and social norms emergence \citep{ashery2025emergent}; to serve as \textit{intervention testbeds} for content moderation \citep{liu2025mosaic} and platform design \citep{touzel2025sandboxsocial}; and, increasingly, to \textit{inform real-world decisions} in policy, public health, and epidemiology \citep{chopra2025limits, phfscience2024digitaltwin}. Simulations are moving from exploring what \textit{might} happen to claiming \textit{why} it happens, \textit{when} it will happen, and what should happen next.

\textbf{A Validation Gap: Realism Without Robustness.}
As the stakes have risen, so has the field's attention to validation \citep{pmlr-v267-anthis25a,larooij2025validation, puelmaTouzel2026validationgap}. However, existing efforts focus almost entirely on whether a simulation \textit{resembles} the world: whether agents behave plausibly \citep{park2023generativeagents}, whether outcomes match empirical patterns \citep{gu2025echochamber, larooij2025validation, yang2024oasis}, whether interaction is grounded in established theory \citep{mou2024hisim, liu2024skepticism}, or whether human evaluators find the results convincing \citep{puelmaTouzel2026validationgap}. We argue that \textit{realism}, while necessary, is not sufficient. A simulation should also be \textit{robust}: its conclusions should remain stable across reasonable alternative configurations, just as empirical findings are expected to survive alternative controls, operationalizations, and sensitivity analyses \citep{simonsohn2020specification, steegen2016increasing}. Without robustness testing, we cannot tell whether a finding reflects the social process being modeled or the design choices used to model it.

\textbf{Architecture Sensitivity and the Butterfly Effect.}
Robustness matters because LLM-based agents introduce a level of architectural complexity that traditional rule-based agent-based modeling does not. A single simulation rests on choices about persona format, memory representation, interaction protocols, and so on \citep{zhou2025pimmur}. Some choices may look interchangeable to researchers, but they are not necessarily equivalent to an LLM \citep{salinas2024butterfly, sclar2024quantifying}. Small differences at the micro level can change how agents interpret context and make decisions, altering local interactions. Through repeated interaction, these local differences can compound into substantially different macro-level outcomes \citep{bertolotti2020sensitivity}. We refer to this cascade as a \textbf{butterfly effect} \citep{ferrara2024butterfly} in the context of LLM social simulations, with Figure \ref{fig:butterfly_effect_illustration} demonstrating an example. 
In Section \ref{sec:case_study}, we demonstrate empirical evidence of the existence of this butterfly effect using two case studies. 
The key finding is that robustness is unevenly distributed across architectural dimensions, and one cannot tell which dimensions matter without auditing them.

In this paper, we argue that \textbf{the scientific claims drawn from LLM-based social simulations should be calibrated to the robustness audits that support them: stronger claims demand stronger audits across architectural design choices, and claims that no audit can support should not be drawn at all.}
Specifically, we make two main contributions: \textit{(1) Demonstrating the butterfly effect in LLM social simulations.} Through two controlled case studies (Prisoner's Dilemma and social network echo chamber simulations) across four LLMs, we show that small architectural perturbations can substantially shift macro-level outcomes, unevenly across both design dimensions and model families; \textit{(2) Introducing a taxonomy and prioritization framework for robustness audits.} We propose \textbf{TRAILS} (\textit{Taxonomy for Robustness Audits In LLM Simulations}), which organizes audit dimensions across agent, interaction, and system levels, and three prioritization heuristics for calibrating audit scope to claim type, simulation complexity, and domain stakes. Our contributions are designed to raise the evidentiary bar for LLM social simulations. As these silicon societies are increasingly proposed as scientific tools and policy testbeds \citep{puelmaTouzel2026validationgap}, robustness audits are needed to ensure their findings reflect stable social dynamics rather than artifacts of simulation design.

\section{Opportunities and Validation Gap in LLM-based Social Simulations}
\label{sec:validation_gap}


\textbf{Scope.}
Building on \textit{Silicon Societies} \citep{puelmaTouzel2026validationgap} and the concept of generative social simulation \citep{pmlr-v267-anthis25a,larooij2025validation}, we focus on LLM-based simulations in which individual human-like agents generate behavior, interact within a shared social environment, and produce collective human outcomes. In this sense, \textit{social} refers to interdependent interaction among agents, while \textit{collective} refers to aggregate patterns emerging from those interactions. We define this scope through a three-level chain: \textbf{micro-level} human-like agents whose reasoning, memory, attitudes, decisions, and communication are generated or mediated by LLMs \citep{yu2024large}; \textbf{meso-level} social interaction among agents within shared environments and institutional rules; and \textbf{macro-level} collective phenomena. To focus on collective human behavior, we exclude three classes of work and the details are provided in Appendix~\ref{appendix:extended-related-works}.

Within this scope, we distinguish two scenarios. In \textbf{goal- or incentive-structured scenarios}, agents act under well-specified incentives, constraints, rules, or payoffs, producing collective outcomes through strategic aggregation, as in negotiation and cooperation \citep{abdelnabi2023negotiation,akata2025repeatedgames,jiang2025cooperation}, population decision-making \citep{mi2025mfllm}, financial market dynamics \citep{yang2025twinmarket}, or macroeconomic activity \citep{li2024econagent}. In \textbf{open-ended scenarios}, agents do not share a single task or payoff structure. Macro-level patterns emerge from micro-level agent behavior and meso-level interaction, such as polarization and echo chambers \citep{gu2025echochamber,piao2025emergence,wang2025decoding,yang2024oasis}, conformity and herd effects \citep{wang2025userbehavior, yang2024oasis}, emergent coordination \citep{orlando2026emergent}, and social norm formation \citep{ren2024socialnorms}. The key distinction is whether collective outcomes arise through \textit{strategic aggregation} in structured decision settings or through \textit{social emergence} in less structured social environments.

\textbf{Beyond Realism: The Missing Robustness Standard.}
Agent-based modeling (ABM) has long been used to study how macro-level social patterns emerge from micro-level agents and local interactions, but it has faced challenges in behavioral realism, empirical validation, replication, and sensitivity to modeling choices \citep{bertolotti2020sensitivity,macy2002factors,ormerod2009validation,wilensky2007making,windrum2007empirical}. By enabling flexible natural-language reasoning, memory, communication, and adaptation, LLMs have made generative agents a promising approach to social simulation, seemingly addressing the challenge of limited behavioral realism \citep{li2024econagent,liu2025mosaic,mou2024hisim, park2023generativeagents,touzel2025sandboxsocial,yang2024oasis}. However, as \citet{larooij2025validation} argue, generative social simulations may increase behavioral realism while also introducing new sources of uncertainty, including black-box model behavior, cultural biases, stochastic variation, alignment effects, and prompt sensitivity. Existing studies have begun to improve the mechanistic validity of LLM social simulations through real-world grounding \citep{gu2025echochamber,liu2025mosaic,wang2025userbehavior}, empirical comparison \citep{gu2025echochamber,yang2024oasis}, theory-driven design \citep{liu2024skepticism, mou2024hisim}, and human evaluation \citep{park2023generativeagents,touzel2025sandboxsocial}. Yet robustness and sensitivity analysis remain underdeveloped. Few studies test whether the same collective outcome survives reasonable perturbations in design choices, such as prompt wording, memory representation, or interaction protocols.

This gap is consequential because LLM agents introduce additional complexity into social simulation. Small implementation choices may propagate through repeated agent interactions and produce large differences in macro-level outcomes. In this sense, LLM social simulations may exhibit a butterfly effect similar to other complex systems \citep{bertolotti2020sensitivity}. This butterfly effect creates risks for both \textit{stability} and \textit{controllability}. Simulations may fail to reproduce the same macro-level pattern under reasonable perturbations, and interventions may fail to shift outcomes in reliable and interpretable ways. Without robustness checks, simulated collective outcomes may reflect arbitrary choices of model, prompt, agent profile, or interaction protocol rather than stable evidence about the underlying social process. This gap motivates expanding realism-centered validation to include systematic robustness audits for LLM social simulation.

\section{Position: Calibrate Robustness Audits to the Strength of Claims}
\label{sec:position}


\subsection{Butterfly Effects and Their Sources in LLM Social Simulation}

\paragraph{Butterfly effects in individual LLMs.} Following the three levels defined in Section~\ref{sec:validation_gap}, we locate robustness risks in LLM social simulations at the levels of individual LLM agents, interaction among agents, and collective outcomes. First, at the micro-level, individual LLMs are already known to be sensitive to small changes in how a task is represented. Minor prompt perturbations, formatting choices, output constraints, and option ordering can change model performance or decisions \citep{pezeshkpour2024sensitivity,razavi2025benchmarking,salinas2024butterfly,sclar2024quantifying}. This matters for social simulation because LLM agents' outputs become actions, messages, memories, and inputs for other agents. Small representational changes at the individual-level can propagate through repeated multi-agent interaction and shape collective outcomes.

\paragraph{Potential propagation of butterfly effects to collective outcomes.}

LLM agents add behavioral complexity to social simulation by making agents more expressive, context-sensitive, and capable of natural-language interaction \citep{pmlr-v267-anthis25a,larooij2025validation}. Prior studies establish sensitivity at the individual-model level \citep{pezeshkpour2024sensitivity,salinas2024butterfly,sclar2024quantifying}. In collective settings, this sensitivity suggests a possible propagation mechanism: a small change in one agent's response may become a different action, message, memory update, or social signal for other agents. These local differences could propagate through interaction protocols and accumulate into large differences in macro-level outcomes. The butterfly metaphor has been used previously to describe how small perturbations in AI systems propagate into disproportionate downstream effects on outcomes such as bias and fairness \citep{ferrara2024butterfly}. Here we extend this framing from single-model pipelines to multi-agent social simulations, where amplification could operate through repeated interaction among LLM agents rather than through a single inference path. Compared with individual-level LLM sensitivity, collective-level butterfly effects remain underexplored, so we illustrate this problem with controlled case studies in Section~\ref{sec:case_study}.

Additionally, LLM social simulations can be unstable in two ways when modeling collective behavior. \textbf{Design-level perturbations} change the substantive simulation setup, such as the model, network topology, or interaction protocol. They test whether a finding is robust across reasonable alternative designs. \textbf{Representation-level perturbations} keep the setup fixed but change how it is presented to the LLM, such as rewriting a persona as prose versus bullet points, reordering equivalent instructions, changing synonymous labels, or formatting memory differently. Both forms of robustness remain underexamined. Representation-level robustness is especially easy to overlook because most studies assume that equivalent textual representations will behave equivalently in LLM social simulations.





\textbf{Claim type and audit strength.} The strength of robustness evidence required scales first with the type of claim a simulation is used to support. We distinguish three levels. \textbf{Exploratory probes} use a simulation to generate hypotheses, illustrate a possible dynamic, or surface patterns worth further study; these claims require modest robustness evidence, but should still show that the result is not an artifact of a single prompt, seed, or model. \textbf{Mechanism claims} use a simulation to argue \textit{why} a collective outcome emerges, such as whether polarization is driven by homophily, selective exposure, moderation design, or memory effects; these claims require stronger checks across agent specification, memory, interaction protocol, environment design, and measurement. \textbf{Policy and intervention claims} use a simulation to evaluate moderation strategies, platform interventions, public-communication strategies, or policy design; these require the highest audit standard, because simulated effects may be misread as reliable predictions about real-world consequences.

\textbf{Simulation complexity and domain stakes.} Beyond claim type, the simulation itself shapes the audit required. The severity of the butterfly effect varies across simulations. Building on social simulation scenarios in Section~\ref{sec:validation_gap}, we distinguish three types of collective-behavior simulations. In \textbf{goal-structured simulations}, agents act in relatively well-specified decision environments with incentives, constraints, rules, or payoffs \citep{akata2025repeatedgames, di2023recognition, fontana2025nicer}. In \textbf{theory-guided open-ended simulations}, agents do not share a single task or payoff structure, but their interactions are structured by social theory, such as opinion dynamics or network theory \citep{gu2025echochamber, piao2025emergence, wang2025decoding}. In \textbf{emergence-driven open-ended simulations}, macro-level patterns arise from less constrained interaction through communication, memory, platform feedback, and network dynamics \citep{park2023generativeagents}. The more open-ended and less theoretically constrained the simulation is, the more challenging the robustness audit becomes.

Domain stakes also shape the required level of audits. \textbf{Lower-stake} simulations involve exploratory demonstrations or low-risk hypothesis generation, while \textbf{higher-stake} simulations involve domains, like public health, policy making, platform governance, or financial markets. LLM social simulations in high-stakes settings are especially valuable, but they require stronger robustness evidence, as their results may inform claims about real-world conditions, institutions, or interventions \citep{saltelli2008global,sargent2010verification}. Together, claim type, simulation complexity, and domain stakes determine how much robustness evidence a simulation result needs before it can support a particular claim: the more ambitious the claim, the more open-ended the simulation, and the higher the stakes, the stronger the audit.



\subsection{Stop Drawing Scientific Claims Beyond What Robustness Audits Support}

LLM-based social simulations are becoming increasingly important tools for studying collective behavior, testing social mechanisms, and evaluating potential interventions \citep{pmlr-v267-anthis25a}. In light of this trend, we argue that \textbf{scientific claims drawn from LLM social simulations should be calibrated to the robustness audits that support them: the stronger the claim---from exploratory probe, to mechanism evidence, to policy guidance---the stronger the audit it requires.} At present, the field lacks shared standards for what level of robustness is required before a simulated outcome can support a scientific claim at each of these levels. This is problematic because of the butterfly effects in LLM social simulation. Under different implementations, the same simulation may support, weaken, or reverse the same claim. For example, a study might use an LLM social simulation to test whether echo chambers emerge among users with different ideologies. The macro-level pattern may appear similar across runs, such as the formation of ideologically clustered communities. However, the underlying process may be unstable. A substantial share of LLM agents may move across different clusters in different runs, or the same ideological group may show different in-group and out-group interaction patterns under slightly different implementations. If these patterns change with small perturbations, then the aggregate echo chamber result is not yet reliable evidence for the claimed mechanism. A simulation may reproduce a plausible macro-level pattern while remaining unstable in the micro- and meso-level process that generates it. 

Thus, we argue for expanding realism-centered validation with robustness audits. Before drawing scientific claims from simulated collective behavior, researchers should ask which design choices were varied, which results remained stable, which results were sensitive, and which assumptions were left unaudited. In Section~\ref{sec:case_study} we show empirically that this calibration is needed: small architectural choices can substantially shift macro-level outcomes, and unevenly across model families. In Section~\ref{sec:taxonomy} we then propose a practical taxonomy for identifying design-level and representation-level perturbations that researchers can audit.

\section{Are LLM Social Simulations Robust? Case Studies of the Butterfly Effect}
\label{sec:case_study}

To test whether LLM-based social simulations are sensitive to seemingly small design choices, we use two controlled case studies that span the simulation types defined in Section \ref{sec:position}. The first is a goal-structured setting: a repeated Prisoner's Dilemma, where agents make decisions under explicit payoffs and a clearly defined action space. This setting is widely used in recent LLM-agent research to study cooperation, reasoning, and strategic behavior \citep{akata2025repeatedgames, brookins2024playing, fontana2025nicer, horton2023large, li2025spontaneous, willis2025will}, and serves as a canonical testbed for incentive-structured collective behavior. The second is an open-ended setting: a social-media echo chamber simulation on a fixed network of agents. This setting reflects a large body of LLM-agent work on polarization, opinion dynamics, and platform design \citep{chuang2023wisdom, ferraro2024agent, gu2025echochamber, piao2025emergence, tornberg2023simulating, wang2025decoding, zheng2024simulating}. By studying both a structured game and an open-ended social media environment, we test whether architecture sensitivity is confined to one kind of simulation or appears across different regimes.

Across both case studies, we use \texttt{gpt-5.2} as the primary model, repeat each condition with $N=30$ independent simulation runs using distinct random seeds, and use each run as the unit of analysis. We report results for \texttt{gpt-5.2} in the main text; full configurations, metric definitions, and statistical procedures are provided in Appendix~\ref{appendix:simulation-details}, and cross-model robustness checks with \texttt{claude-haiku-4-5}, \texttt{gemini-2.5-flash}, and \texttt{deepseek-v3.2} in Appendix~\ref{appendix:extended-results}. The three strategic persona formats are reproduced in Figure~\ref{fig:persona_format_panel}. Our results show that these cross-model results reveal substantial heterogeneity rather than uniform replication: the magnitude of architecture sensitivity is itself model-dependent, which we treat as a second axis of fragility throughout the paper.

\subsection{Goal-Structured Simulation: Repeated Prisoner's Dilemma}
\label{sec:PD-case-study}
 
Two players play a repeated Prisoner's Dilemma for $T=10$ rounds. In each round, each player simultaneously chooses Cooperate ($C$) or Defect ($D$), with the canonical payoff structure $(C,C)\to(3,3)$, $(C,D)\to(0,5)$, $(D,C)\to(5,0)$, and $(D,D)\to(1,1)$. Following common practice in previous studies \citep{akata2025repeatedgames, di2023recognition, fontana2025nicer}, we run both single-agent play against fixed benchmark policies and two-agent play between LLM-controlled agents. We evaluate agents using average payoff and cooperation rate, where cooperation rate serves as the measure of prosocial tendency. Detailed configurations and metric definitions are provided in Appendix \ref{appendix:simulation-details}. We test three perturbations, each targeting a design choice that is inconsistently reported and rarely ablated across the LLM-cooperation literature: persona format, game instruction framing, and memory representation. 



\begin{figure}[t]
    \centering
    \includegraphics[width=\textwidth]{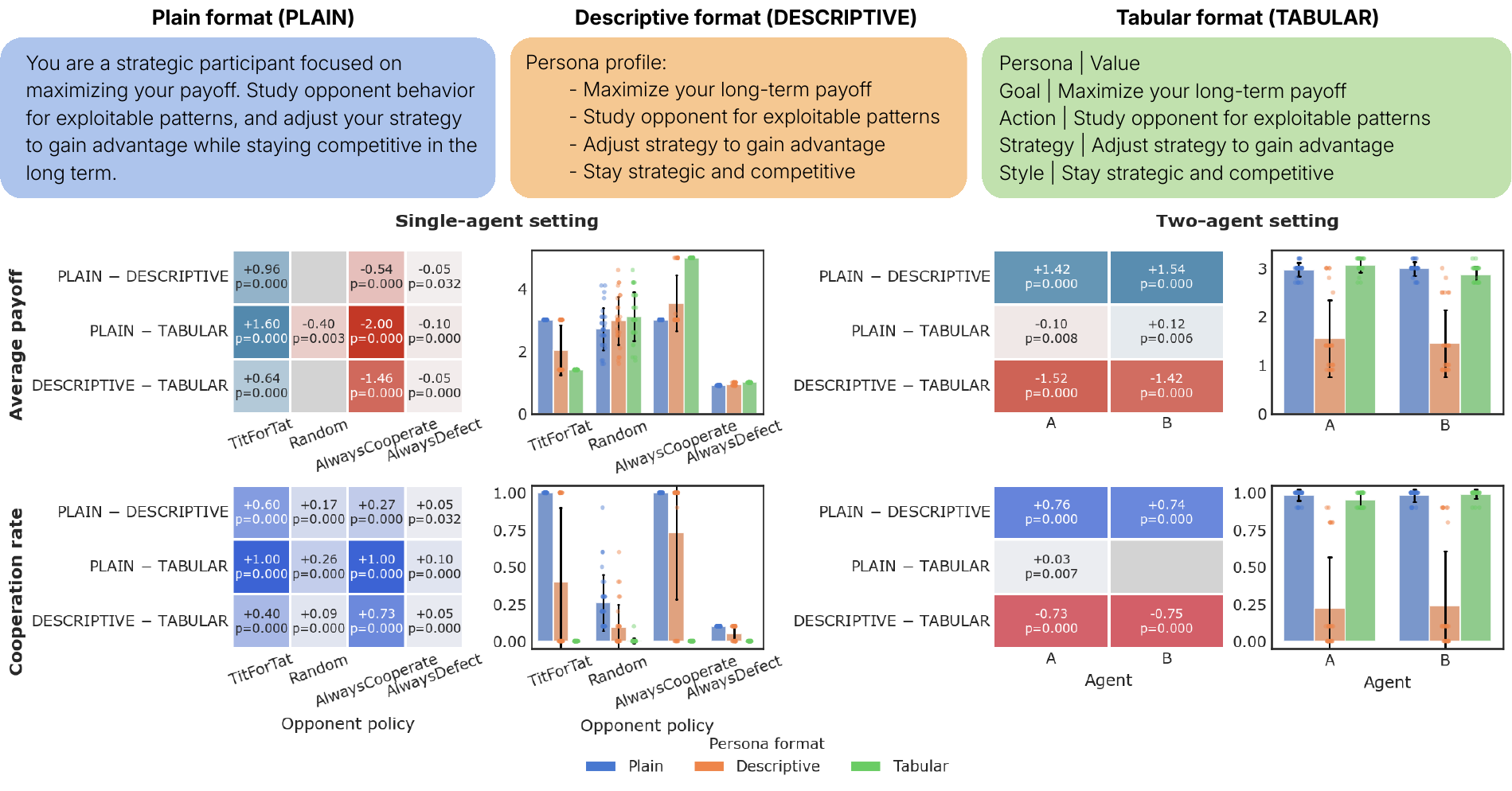}
    \caption{
    \textbf{Effect of persona format on Prisoner's Dilemma outcomes.}
    Results are shown for the single-agent setting (left) and two-agent setting (right). 
    Heatmaps report statistically significant pairwise differences in average payoff and cooperation rate across persona formats ($p < .05$, two-sided Mann--Whitney U test); gray cells indicate non-significant comparisons. 
    Bar plots show run-level distributions, with error bars denoting one standard deviation. 
    Bar colors denote persona format: \textcolor{blue}{\textsc{plain}} (blue), \textcolor{orange}{\textsc{descriptive}} (orange), and \textcolor{green}{\textsc{tabular}} (green).
    }
    \label{fig:persona_format_panel}
    \vspace{-1em}
\end{figure}

\textbf{Perturbation 1: Persona format.}
We first test whether surface-level formatting of a persona prompt can shift simulation outcomes independently of its content. Holding the underlying meaning fixed---the agent is strategic and aims to maximize its payoff---we vary only the presentation across three formats: a paragraph of plain prose (\textsc{Plain}), a descriptive bullet list (\textsc{Descriptive}), and a structured key--value table (\textsc{Tabular}). As illustrated in Figure \ref{fig:persona_format_panel}, changing only the persona format produces statistically significant effects on both payoff and cooperation. In the single-agent setting, under the \textit{AlwaysCooperate} opponent policy, \textsc{Tabular} personas earn 1.46 more points per round than \textsc{Descriptive} personas on average, on a 0--5 payoff scale (95\% CI: [1.14, 1.78], $d=2.30$, $p<0.001$), and 2.00 more points per round than \textsc{plain} personas ($p<0.001$). These payoff gains correspond to lower cooperation: across all four fixed opponent policies, \textsc{Tabular} personas are consistently the least cooperative. In the two-agent setting, where both agents use the same persona format, the effect is especially large: \textsc{Descriptive} personas cooperate 76 percentage points less often than \textsc{Plain} personas ($p<0.001$), and 73 percentage points less often than \textsc{Tabular} personas ($p<0.001$). These results show that semantically equivalent personas can shift the apparent equilibrium of the simulation.


\textbf{Perturbation 2: Game-instruction framing.}
We next hold the payoff matrix and persona fixed, using \textsc{plain} as the default persona, and vary only the framing of the game instructions. We test three framings: canonical game-theory framing (\textsc{Canonical}), moralized framing (\textsc{Moralized}), and risk framing (\textsc{Risk}). Overall, game-instruction framing significantly changes both cooperation and payoff. The most consistent pattern is that \textsc{Moralized} framing increases cooperation relative to \textsc{Canonical} and \textsc{Risk} framing. Especially in the two-agent setting, \textsc{moralized} framing also yields higher average payoff, suggesting that a small change in how the same payoff matrix is described can shift agents to a more cooperative equilibrium. Full instruction text and results are in Appendix \ref{appendix:extended-results}.
 
\textbf{Perturbation 3: Memory representation.}
Finally, we hold the persona fixed to \textsc{Plain} and the game framing fixed to \textsc{Canonical}, and vary only how the interaction history is shown to the agent. Unlike persona format and game-instruction framing, memory representation has only small effects on the main outcomes: most payoff shifts are below 0.2 on a 0--5 scale, and most cooperation-rate shifts are below 0.1 on a 0--1 scale. Thus, changing how the same history is represented can affect behavior, but it does not shift the simulation equilibrium in the way persona format or game framing does. Full memory text and statistical results are shown in Appendix \ref{appendix:extended-results}.

\subsection{Open-Ended Simulation: Echo Chamber on a Social Network}
\label{sec:EC-case-study}

\begin{figure}[t]
    \centering
    \includegraphics[width=\textwidth]{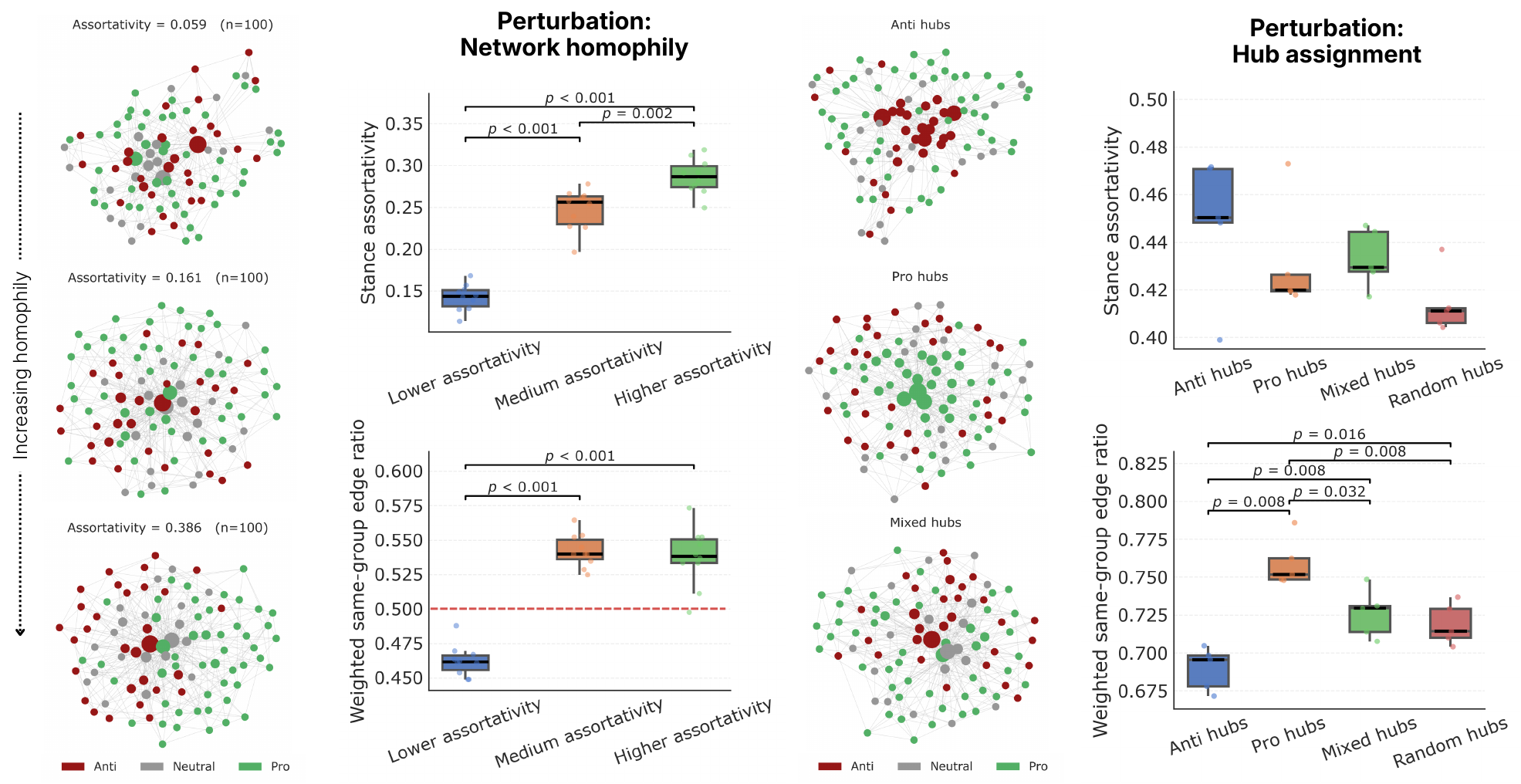}
    \caption{\textbf{Effects of network homophily and hub assignment on echo-chamber outcomes.}
    Left panels show example input networks with increasing stance homophily, operationalized as higher initial network assortativity. Right panels show example networks with the same degree sequence but different hub assignments, where high-degree nodes are occupied by anti-stance, pro-stance, mixed, or randomly assigned agents. Boxplots report two run-level metrics computed on the simulated interaction network: stance assortativity and weighted same-group edge ratio. Significant pairwise differences are annotated using two-sided Mann--Whitney U tests with $p < 0.05$.}
    \label{fig:homophily_hub_panel}
    \vspace{-1em}
\end{figure}

The echo-chamber simulation consists of 100 agents with distinct personas. Each persona includes a short bio and a stance on whether advanced AI systems should be regulated, represented on a five-point scale from strongly against regulation to strongly support regulation. Each agent's stance is \emph{frozen} throughout the run, allowing us to isolate structural and interaction effects from belief-update dynamics. This scope choice means our perturbations test sensitivity in the formation of \emph{structural} echo chambers rather than in \emph{opinion} polarization driven by belief updating; whether the same perturbations matter for the latter is an important question for future work. Agents interact for 15 rounds on a fixed power-law-like network with a fixed degree sequence, and can only see posts from their direct neighbors. In each round, active agents select one of four actions: \textsc{Post}, \textsc{Repost}, \textsc{Reply}, or \textsc{Do Nothing}, conditioned on their persona, stance, and recent neighbor posts.

We measure two echo-chamber and polarization metrics on the interaction network \citep{gu2025echochamber, wang2025decoding, piao2025emergence}: stance assortativity and weighted same-group edge ratio. We test five perturbations that target common but often under-audited design choices in LLM echo-chamber simulations: input-network homophily, hub assignment, activation probability, memory window, and recommendation feed size. These choices span structural conditions that shape who can interact with whom and interaction-level conditions that shape who becomes active, what content agents see, and what recent context conditions their actions. Detailed definitions and configurations are provided in Appendix \ref{appendix:simulation-details}.

We deliberately vary homophily and hub assignment as \emph{context-relevant} structural conditions, not as equivalent representations of a fixed social system. Because these conditions can substantively affect echo-chamber formation, sensitivity is expected and is not, by itself, evidence of model fragility. These contrasts instead test whether the claimed phenomenon persists across structural conditions or must be scoped to a subset of them, following the claim-level interpretation formalized in the Social Simulation Audit Card (Section~\ref{sec:taxonomy}).

\textbf{Perturbation 1: Initial network homophily.}
As shown in Figure \ref{fig:homophily_hub_panel}, we compare three input networks with only modest increases in initial stance assortativity. Despite these small structural shifts, final interaction-network stance assortativity increases substantially, from $0.142$ to $0.247$ and $0.287$, with all pairwise differences significant (Mann--Whitney $p \leq 0.002$). The weighted same-group edge ratio shows an even clearer threshold pattern: the lowest-homophily network remains below $0.5$ ($M=0.462$, 95\% CI $[0.454,0.470]$), meaning weighted interactions are still mostly cross-stance, whereas the two higher-homophily networks are clearly above $0.5$ ($M=0.542$, 95\% CI $[0.534,0.551]$; $M=0.538$, 95\% CI $[0.522,0.553]$), indicating majority within-stance interaction. Thus, a modest increase in initial network assortativity is enough to shift the simulation from mostly cross-group exposure to echo-chamber-like interaction.

\textbf{Perturbation 2: Hub assignment.}
We next vary whether the highest-degree nodes are assigned to anti-regulation, pro-regulation, mixed, or random agents, while keeping the degree sequence fixed. Hub assignment does not produce significant differences in final stance assortativity. However, all weighted same-group edge ratios are clearly above $0.5$, indicating echo-chamber-like interaction in every hub condition. The effect is strongest when pro-regulation agents occupy the hubs ($M=0.759$, 95\% CI $[0.739,0.779]$) and weakest when anti-regulation agents occupy the hubs ($M=0.690$, 95\% CI $[0.672,0.707]$), with several pairwise differences significant. This suggests that hub assignment changes the strength, though not the presence, of echo-chamber effects.

\textbf{Perturbations 3--5: Activation probability, memory window, and recommendation feed size.}
We next vary three interaction-level parameters: activation probability, memory window, and recommendation feed size. Activation probability and memory window do not produce significant changes in either stance assortativity or weighted same-group edge ratio. In contrast, increasing the recommendation feed size from $5$ to $10$ posts strengthens echo-chamber outcomes: stance assortativity increases from $0.247$ to $0.277$ ($p=0.014$), and the weighted same-group edge ratio increases from $0.542$ to $0.568$ ($p=0.003$). Thus, among these three perturbations, feed size is the main factor that amplifies echo-chamber-like interaction. Additional figures are presented in Appendix \ref{appendix:extended-results}.

\textbf{Finding Summary.}
Across both case studies, LLM social simulations are neither uniformly fragile nor uniformly robust. Sensitivity is uneven along two axes. First, \emph{across architectural dimensions}, and unevenly within that axis: persona format can produce equilibrium-flipping shifts in macro outcomes (up to 76 pp in cooperation rate); game-instruction framing, network homophily, hub assignment, and recommendation feed size produce smaller but consistent and statistically significant shifts; while memory representation, memory window size, and activation probability have small or null effects. Second, \emph{across model families}: the same persona-format perturbation that produces a 76-percentage-point cooperation gap in \texttt{gpt-5.2} produces a comparably large gap in \texttt{claude-haiku-4-5} (${\sim}77$ pp), a moderate gap in \texttt{gemini-2.5-flash} (${\sim}36$ pp), and essentially no effect in \texttt{deepseek-v3.2} (${\sim}1$ pp; Appendix \ref{appendix:extended-results}). Model identity is therefore itself a robustness dimension: cross-model checks on a small set of models can leave fragility undetected. The key question is not whether every simulation is fragile, but which design choices must be audited---and across which models---before its results can be trusted. What should researchers perturb? Which dimensions matter for which kinds of claims? In the next section, we introduce a taxonomy for answering these questions systematically.

\section{Toward Robustness Audits for LLM Social Simulations}
\label{sec:taxonomy}




To identify actionable perturbations, we synthesize common design choices in existing work on LLM social simulations for collective behavior and propose \textbf{TRAILS} (\textit{Taxonomy for Robustness Audits In LLM Simulations}), a practical framework for auditing whether simulated collective outcomes are robust to changes in simulation systems (Table~\ref{tab:trails}). TRAILS comprises \textbf{TRAILS-D} (design-level perturbations) and \textbf{TRAILS-R} (representation-level perturbations). Measurement and evaluation sensitivity remain important, but we treat them as part of the downstream evaluation pipeline rather than as perturbations of the simulation.

\begin{table}[t]
\caption{TRAILS: Taxonomy for Robustness Audits In LLM Simulations.}
\label{tab:trails}
\centering
\scriptsize
\setlength{\tabcolsep}{3pt}
\renewcommand{\arraystretch}{1.05}
\begin{tabular}{p{0.08\linewidth} p{0.20\linewidth} p{0.56\linewidth} p{0.08\linewidth}}
\toprule
\textbf{Level} & \textbf{Dimension} & \textbf{Description} & \textbf{Examples} \\
\midrule

\multicolumn{4}{l}{\textbf{TRAILS-D: Design-level perturbations}} \\
\midrule

\textbf{Micro}
& Model substrate
& Which LLM and inference settings underlie agent behavior.
& \citep{zhou2024sotopia} \\

& Agent specification
& How agents are represented as social actors.
& \citep{park2023generativeagents,li2025llm} \\

& Internal state and cognition
& How agent beliefs, attitudes, goals, emotions, and reasoning are represented.
& \citep{liu2024skepticism,wang2023humanoid}\\

& Memory and temporality
& How past events are stored, summarized, retrieved, forgotten, and reflected upon.
& \citep{park2023generativeagents,wang2025userbehavior} \\

\midrule

\textbf{Meso}
& Interaction protocol
& Who interacts, when, under what visibility rules, and through which actions.
& \citep{mou2024hisim,yang2024oasis} \\

& Intervention design
& How interventions are selected, delivered, timed, targeted, and framed.
& \citep{liu2025mosaic,liu2024skepticism} \\

\midrule

\textbf{Macro}
& Environment structure
& How spatial, institutional, or network structures shape collective outcomes.
& \citep{yang2024oasis,gu2025echochamber,piao2025agentsociety} \\

& Population and scale
& How population composition, size, and heterogeneity shape collective outcomes.
& \citep{yang2024oasis,wang2025yulanonesim,piao2025agentsociety}\\

\midrule

\multicolumn{4}{l}{\textbf{TRAILS-R: Representation-level perturbations}} \\
\midrule

&
Representational format
& How equivalent information is structurally formatted or encoded.
& \citep{salinas2024butterfly,sclar2024quantifying} \\

&
Instruction hierarchy
& How equivalent instructions are ordered, placed, or assigned authority.
& \citep{salinas2024butterfly,sclar2024quantifying} \\

&
Linguistic framing
& How equivalent meanings are expressed through wording.
& \citep{salinas2024butterfly,sclar2024quantifying} \\

&
Context representation
& How interaction history, memory, and metadata are represented or compressed.
& \citep{park2023generativeagents,salinas2024butterfly}\\

&
Interaction sequencing
& How the same interaction is initialized, ordered, or terminated.
& \citep{mou2024hisim} \\

\bottomrule
\end{tabular}
\end{table}

\textbf{TRAILS-D} summarizes design-level perturbations that may change the substantive setup of an LLM-based social simulation. Following the three-level structure of LLM social simulation defined in Section~\ref{sec:position}, we organize these perturbations into the eight dimensions shown in Table~\ref{tab:trails}. These perturbations matter because collective outcomes are produced by the coupling between LLM agents and simulation design. A finding about a macro-level social phenomenon may therefore depend on design choices that are not part of the claimed social mechanism. TRAILS-D asks whether the same substantive claim survives reasonable alternatives in these design choices. We provide some examples of such perturbations in Table~\ref{tab:trails_d}. \textbf{TRAILS-R} focuses on perturbations that keep the substantive simulation condition fixed but change how that condition is represented to LLMs. This matters because LLM agents receive the simulation through text and interface structure, and prior work shows that small prompt variations can substantially affect LLM outputs \citep{salinas2024butterfly,sclar2024quantifying}. Formatting, instruction order, labels, context compression, and interaction sequencing can affect how agents interpret a situation and produce behavior. TRAILS-R therefore audits representation and interface sensitivity by asking whether the same collective outcome survives these representation-level changes. If it does not, the finding should be reported as interface-sensitive rather than treated as robust evidence about the social process. We provide some examples of representation-level perturbations in Table~\ref{tab:trails_r}.

\textbf{Prioritizing audits.} No single study can audit every TRAILS dimension, and we do not suggest that every paper should. We propose three heuristics for deciding which dimensions to perturb. First, \emph{align audits with the claim's mechanism}: a claim that cooperation arises as a stable strategic equilibrium requires auditing persona format and game framing more than memory representation, because the former shape strategic disposition while the latter shape only how history is recalled. Second, \emph{ensure each perturbation spans a meaningful range}: two prompt variants are weaker evidence than a sweep across several reasonable alternatives along the same axis. Third, \emph{audit across model families before claiming generality}: as Section~\ref{sec:case_study} shows, the same perturbation can produce dramatic effects in one model and essentially none in another, so claims phrased about ``LLM agents'' in general require evidence from multiple frontier models. Together with the claim-type hierarchy in Section~\ref{sec:position}, these heuristics let researchers calibrate audit scope to the strength of the claim being made.

\noindent \textbf{The Social Simulation Audit Card.}
We introduce the \textit{Social Simulation Audit Card} to operationalize TRAILS for a stated scientific claim. The card guides researchers through five steps: stating the claim, mapping perturbations to TRAILS and context relevance, registering perturbations and preservation criteria, reporting sensitivity, and defining the claim's scope. Context relevance is claim-specific: context-relevant perturbations represent substantively meaningful conditions of the modeled phenomenon, whereas context-irrelevant perturbations probe implementation dependence. For example, network homophily and hub assignment are context-relevant structural conditions in the echo-chamber simulation. Accordingly, context-relevant sensitivity helps scope where a claim holds, while substantial sensitivity to context-irrelevant choices indicates implementation dependence. Robustness therefore concerns preservation of a stated claim within its specified scope, not numerical invariance across substantively different social conditions. Outcomes are classified as \textit{Robust}, \textit{Scope-limited}, \textit{Implementation-sensitive}, or \textit{Inconclusive}, bounded by the audited conditions. Appendix~\ref{appendix:audit-card} summarizes the procedure and its application to the reported case studies (Table~\ref{tab:case-study-perturbations}).

\section{Alternative Views and Discussion}

\textbf{Alternative view 1: All empirical methods exhibit sensitivity.}
One objection is that LLM simulations are not unique: Markov chain Monte Carlo estimates vary with seeds, agent-based models depend on initialization and parameters, and human-subject studies are sensitive to sampling, measurement, and analytic choices. We agree, but the fact that every other empirical tradition has developed protocols for handling its sensitivities is precisely why LLM-based social simulations need one of their own. Machine learning uses train--test splits, cross-validation, ablations, stress tests, and benchmarks; social science uses construct-validity checks, multiverse analysis \citep{steegen2016increasing}, and specification curve analysis \citep{simonsohn2020specification}. LLM simulations introduce a distinct perturbation surface: textual, high-dimensional, and semantically opaque \citep{mizrahi2024state, sclar2024quantifying}. TRAILS extends this validation logic by asking whether a simulation finding is stable enough to be worth validating against reality in the first place.

\textbf{Alternative view 2: LLM social simulations should be limited to hypothesis generation.}
A stronger objection is that LLM agents are not human, so robustness cannot make their outputs evidence about human social behavior. We take this seriously. The position we advance in Section~\ref{sec:position} is already calibrated to this concern: audit strength scales with claim type, simulation complexity, and domain stakes. The field is already using LLM simulations as policy testbeds for platform interventions \citep{huang2026policysim}, prosocial-behavior policy evaluation \citep{zhou2026investigating}, public-health interventions \citep{hou2025can}, and epidemic decision-making \citep{aoki2026ai}---uses that fall at the policy-claim level of our framework and therefore demand the strongest audits. Claims that survive TRAILS-D and TRAILS-R perturbations across frontier models deserve more weight than those that do not, even if they remain weaker than human-subject evidence. We invite discussion on where this evidentiary gradient should sit.


\textbf{Alternative view 3: Comprehensive robustness audits are prohibitively expensive.}
A third objection is practical: auditing every TRAILS dimension across multiple models and seeds quickly becomes infeasible, especially for small labs. We acknowledge this, and our framework is graded along two axes that respond to it. First, audit strength scales with claim strength: exploratory probes do not require comprehensive audits. Second, the prioritization heuristics in Section~\ref{sec:taxonomy} let researchers concentrate their compute budget on the dimensions most likely to confound the claim being made. Transparent reporting of which dimensions were tested and which were left unaudited is itself a contribution, since it lets the field aggregate evidence across studies without requiring any single study to be exhaustive.

\textbf{A call for robustness audits in LLM social simulation.} 
LLM-based social simulations are increasingly used to explore collective behavior, test mechanisms, and evaluate interventions. As these systems move toward scientific inference and socially consequential applications, plausible agent behavior and realistic-looking macro outcomes are no longer sufficient. TRAILS provides a starting vocabulary for this standard. To move toward robustness audits in LLM social simulation, we call on the community to pursue two lines of action: \textbf{(1) Improve robustness reporting in future LLM social simulation studies.} Future simulation work should state the simulation scenario, domain stakes, and the evidentiary role of the simulation along the claim-type spectrum introduced in Section~\ref{sec:position}---exploratory probe, mechanism claim, or policy claim---and calibrate robustness audits accordingly. Studies should report which perturbations were tested, which findings remained stable, which findings were sensitive, and which dimensions were left unaudited. We acknowledge that no single study can test every possible perturbation; audits should prioritize the design and representation choices most relevant to the claim; \textbf{(2) Build shared infrastructure for robustness auditing.} The field needs reusable perturbation libraries for LLM social simulation settings. Open-source datasets and benchmarks for robustness audits are needed. It also needs more empirical work that systematically tests which perturbations matter most, when butterfly effects occur, and how micro- or meso-level instability propagates into macro-level outcomes.

\bibliographystyle{plainnat}
\bibliography{references}

\newpage
\appendix

\section*{Version Statement}
This position paper examines how implementation choices affect scientific claims from LLM social simulations, using two case studies to motivate robustness audits and introduce TRAILS. A separate empirical study \citep{ye2026butterflyeffectllmsocial} further provides a more extensive analysis of sensitivity across perturbations and models, interaction amplification and attenuation, and behavioral and collective outcomes.

\section{TRAILS (Taxonomy for Robustness Audits In LLM Simulations)}

\begin{table}[htbp]
\caption{TRAILS-D: Design-level perturbations}
\label{tab:trails_d}
\centering
\footnotesize
\begin{tabular}{p{0.1\linewidth} p{0.24\linewidth} p{0.52\linewidth}}
\toprule
\textbf{Level} & \textbf{Dimension} & \textbf{What to perturb} \\
\midrule

\textbf{Micro} 
& Model substrate 
& Model family, model size, base vs. instruction-tuned models, alignment, temperature, top-$p$, random seed, safeguards. \\

& Agent specification 
& Demographics, ideology, personality, goals, prior beliefs, real-data-grounded vs. synthetic personas, relationship attributes. \\

& Internal state and cognition
& Beliefs, attitudes, emotions, needs, moral values, reasoning style, reflection rules, belief updating, preference updating. \\

& Memory and temporality
& last-$k$ memory, summarized memory, episodic memory, retrieval rules, forgetting, reflection frequency, temporal granularity. \\

\midrule

\textbf{Meso} 
& Interaction protocol 
& Turn order, fixed or free turns, dyadic or group interaction, available actions \\

& Intervention design
& Moderator type, target selection, timing, frequency, rule-based vs. LLM-based intervention. \\

\midrule

\textbf{Macro} 
& Environment structure 
& Network topology, feed ranking, recommendation system, moderation rules, platform affordances, institutional rules. \\

& Population and scale 
& Number of agents, demographic distribution, ideological balance, heterogeneity, real vs. synthetic population construction. \\

\bottomrule
\end{tabular}
\end{table}

\begin{table}[htbp]
\caption{TRAILS-R: Representation-level perturbations}
\label{tab:trails_r}
\centering
\footnotesize
\begin{tabular}{p{0.22\linewidth} p{0.22\linewidth} p{0.48\linewidth}}
\toprule
\textbf{Category} & \textbf{Perturbation} & \textbf{Example} \\
\midrule

\textbf{Representational format} 
& Formatting 
& Narrative prose vs. bullet points; paragraph vs. table; bolded keywords; numbered lists. \\

& Delimiter choice 
& Conversation history enclosed in code blocks, XML tags, JSON, or plain text. \\

& Output schema 
& Free-form paragraph vs. short reply vs. JSON with rationale. \\

\midrule

\textbf{Instruction hierarchy} 
& Instruction order 
& “Update your opinion, then write a reply” vs. “Write a reply, then update your opinion.” \\

& Role/message placement 
& The same instruction placed in the system message vs. the user message. \\

\midrule

\textbf{Linguistic framing} 
& Agent naming 
& Agent A/B vs. Alice/Bob. \\

& Persona wording 
& ``Left'' vs. ``Democrat''; ``Right'' vs. ``Republican.'' \\

& Moderation framing 
& Same intervention framed as a warning, suggestion, community reminder, or bridge statement. \\

& Few-shot examples 
& Same examples framed differently. \\

\midrule

\textbf{Context representation} 
& Memory representation 
& Same history shown as transcript, JSON list, bullet summary. \\

& Context window 
& Full history vs. last three turns; same content with early background removed. \\

& Message metadata 
& Adding timestamps, usernames. \\

\midrule

\textbf{Interaction sequencing} 
& Initial seed turn 
& Same stance, but a slightly different opening statement. \\

& Turn order 
& Which agent speaks first. \\




\bottomrule
\end{tabular}
\end{table}

\newpage
\section{Extended Related Works}
\label{appendix:extended-related-works}

\subsection{From Agent-Based Modeling to Generative Social Simulation}
\label{sec:abm2gabm}

Agent-based modeling (ABM) has provided a bottom-up framework for studying how macro-level social patterns emerge from micro-level agents and local interactions \citep{reynolds1987flocks,schelling1971dynamic,macy2002factors}. However, traditional ABMs have also been criticized for relying on hand-coded behavioral rules and challenges in empirical validation, replication, and sensitivity to modeling choices \citep{windrum2007empirical,ormerod2009validation,wilensky2007making,bertolotti2020sensitivity}. Recent advances in LLMs make generative agents a promising way. The literature on LLM agents and multi-agent systems for social simulation is already broad, ranging from collaborative task teams \citep{qian2023chatdev,hong2023metagpt, zhou2023webarena} to embodied agents \citep{ren2025simworld}, game-like environments \citep{vezhnevets2025multiactor}, user simulations in recommendation system \citep{wang2025userbehavior}, economic simulations \citep{li2024econagent}, and social media simulations \citep{yang2024oasis,touzel2025sandboxsocial,gu2025echochamber}.  In this work, we focus on a narrower scope: LLM-based social simulations for collective human behavior. 

\paragraph{Scope.}
Building on \textit{Silicon Societies} \citep{puelmaTouzel2026validationgap} and the eligibility criteria for generative social simulation proposed by \citet{pmlr-v267-anthis25a,larooij2025validation}, we focus on LLM-based simulations in which individual human-like agents generate behavior, interact within a shared social environment, and produce collective human outcomes. In this sense, \textit{social} refers to interdependent interaction among agents, while \textit{collective} refers to aggregate patterns emerging from those interactions. We define this scope through a three-level chain: \textbf{micro-level} human-like agents whose reasoning, memory, attitudes, decisions, and communication are generated or mediated by LLMs; \textbf{meso-level} social interaction among agents within shared environments and institutional rules; and \textbf{macro-level} collective phenomena such as polarization, echo chambers, norm emergence, information diffusion, and mobilization. 

Additionally, we exclude three classes of work to focus on modeling collective human behavior: single-agent or user simulations without social interaction among agents \citep{ren2024bases}; simulations where agents primarily represent non-human, embodied, or aggregate actors such as robots, states, countries, or game agents \citep{altera2024projectsid,hua2023waragent,mandi2023roco,ren2025simworld}; and multi-agent LLM systems designed primarily for task completion or evaluation rather than modeling human social behavior \citep{chan2023chateval,hong2023metagpt,qian2023chatdev,zhou2023webarena}

\paragraph{Social simulation scenarios.}
Within this scope, we distinguish two broad scenarios. In \textbf{goal- or incentive-structured scenarios}, agents act in relatively well-specified decision environments with explicit or implicit incentives, constraints, rules, or payoffs. Collective outcomes arise through strategic aggregation, as in negotiation and cooperation \citep{abdelnabi2023negotiation,akata2025repeatedgames,jiang2025cooperation}, population decision-making \citep{mi2025mfllm}, financial market dynamics \citep{yang2025twinmarket}, or macroeconomic activity \citep{li2024econagent}. In \textbf{open-ended scenarios}, agents do not share a single explicit task, objective, or payoff structure. Macro-level patterns emerge from micro-level agent behavior and meso-level interaction, as in social media discussion \citep{yang2024oasis,touzel2025sandboxsocial}, opinion dynamics and mobilization \citep{mou2024hisim}, polarization and echo chambers \citep{gu2025echochamber,piao2025emergence,wang2025decoding,yang2024oasis}, information or misinformation spread \citep{liu2024skepticism,liu2025mosaic,park2023generativeagents}, conformity and herd effects \citep{yang2024oasis,wang2025userbehavior}, emergent coordination \citep{park2023generativeagents, orlando2026emergent}, social norm formation \citep{ren2024socialnorms}, and user behavior in digital platforms \citep{wang2025userbehavior}. The key distinction between the two types is whether collective outcomes arise primarily through \textit{strategic aggregation} in structured decision settings or through \textit{social emergence} from less structured social environments.

By allowing agents to reason, remember, communicate, and adapt through natural language, LLM-based agents appear to address the challenge of limited behavioral realism in ABM. Recent work has used LLM agents to simulate emergent coordination \citep{park2023generativeagents}, macroeconomic decision-making \citep{li2024econagent}, social media interaction \citep{yang2024oasis,touzel2025sandboxsocial}, echo chamber formation \citep{gu2025echochamber}, social movement dynamics \citep{mou2024hisim}, fake news attitude dynamics \citep{liu2024skepticism}, and information dissemination and moderation \citep{liu2025mosaic}. These studies illustrate the promise of LLM agents for modeling collective behavior.


However, LLMs do not automatically solve the other challenges of simulation. As \citet{larooij2025validation} argue, generative social simulations may increase behavioral realism, but they also introduce new sources of uncertainty, including black-box model behavior, cultural biases, stochastic variation, alignment effects, and prompt sensitivity. Existing studies have begun to improve the realism, empirical grounding, and mechanistic validity of LLM social simulations through real-world grounding \citep{gu2025echochamber,liu2025mosaic,wang2025userbehavior}, empirical comparison \citep{yang2024oasis,gu2025echochamber}, theory-driven design \citep{mou2024hisim,liu2024skepticism}, and human evaluation \citep{park2023generativeagents,touzel2025sandboxsocial}. Yet robustness and sensitivity analysis remain underdeveloped. Fewer studies test whether the same collective outcome survives reasonable perturbations in design choices, such as prompt wording, agent profiles, memory representation or interaction protocols.

\subsection{Repeated Prisoner's Dilemma as a Goal-Structured Testbed}

The repeated Prisoner's Dilemma is a foundational paradigm for studying the tension between short-term self-interest and long-term mutual benefit, and a recurring testbed for LLM-agent cooperation \citep{akata2025repeatedgames, fontana2025nicer, willis2025will, brookins2024playing, li2025spontaneous}. Recent work has used this paradigm to make claims about whether LLMs are ``cooperative'' or ``selfish,'' whether they reason from history, and how their behavior compares to human play \citep{akata2025repeatedgames, horton2023large}. We use it here as a tightly controlled environment: the action space, payoffs, and horizon are fully specified, leaving architectural and prompt-level choices as the main sources of variation.

We focus on three small perturbations that are common in LLM-agent implementations but rarely audited jointly. First, we vary \textbf{persona format}, because personas are widely used to specify agent goals, traits, and social roles, yet recent work shows that persona-based simulations can introduce systematic validity risks if their construction is not carefully validated \citep{li2025llm}. Our test asks a narrower representation-level question: whether the same persona content behaves differently when written as prose, a descriptive list, or a table. Second, we vary \textbf{game-instruction framing}, because prior work shows that LLM behavior in repeated games can change when the same strategic setting is described differently, for example through robustness checks, payoff-matrix variations, or reasoning prompts \citep{akata2025repeatedgames, lore2024strategic}. Third, we vary \textbf{memory representation}, because repeated-game behavior depends on how agents read and use prior interaction history; existing Prisoner's Dilemma studies explicitly test whether LLMs can parse gameplay logs and condition decisions on historical behavior \citep{fontana2025nicer}. Together, these perturbations test whether claims about cooperation, selfishness, or strategic adaptation survive small changes in how the same game, agent, and history are represented to the model.

\subsection{Echo Chambers as an Open-Ended Social Simulation Testbed}

Echo chambers and polarization are among the most studied collective phenomena in LLM-based social simulation \citep{gu2025echochamber, wang2025decoding, ferraro2024agent, zheng2024simulating, piao2025emergence, tornberg2023simulating, yang2024oasis, chuang2024simulating}. Recent work has used LLM agents on social networks to ask whether echo chambers emerge, what role homophily and recommendation play, and how interventions reshape exposure \citep{gu2025echochamber, tornberg2023simulating, liu2025mosaic, touzel2025sandboxsocial}. We adopt this paradigm as a representative open-ended simulation in which collective patterns emerge from local interaction rather than explicit optimization.

We test five perturbations that target core design choices in open-ended social-network simulations: homophily, hub assignment, activation probability, memory window, and recommendation feed size. These choices are motivated by prior echo-chamber and LLM-agent studies, which show that echo chambers emerge from the joint effects of network structure, selective exposure, memory/context, and engagement dynamics. First, we vary \textbf{input-network homophily}, because prior work treats homophily and structurally clustered networks as central mechanisms through which like-minded users become locally concentrated and echo chambers form \citep{gu2025echochamber, wang2025decoding, ferraro2024agent}. Second, we vary \textbf{hub assignment} while holding the degree sequence fixed, because scale-free social networks contain highly connected users whose placement can disproportionately affect information diffusion and group reinforcement; prior LLM-agent simulations similarly emphasize that scale-free or hub-dominated networks better approximate real social media than fully connected or random interaction settings. Third, we vary \textbf{activation probability}, because social media outcomes depend not only on who is connected to whom, but also on which users become active and contribute content in each round. Fourth, we vary the \textbf{memory window}, since LLM-agent frameworks commonly rely on short-term or retrieved interaction histories to condition agent behavior, but the amount of remembered context is often treated as an implementation detail rather than an audited design choice. Finally, we vary \textbf{recommendation feed size}, because prior work shows that selective exposure and recommendation mechanisms can increase engagement, reduce cross-group interaction, and strengthen echo-chamber formation \citep{wang2025decoding, ferraro2024agent}. Together, these perturbations span two levels of the simulation architecture: structural opportunity conditions, through homophily and hub assignment, and interaction-level exposure conditions, through activation, memory, and feed size. This allows us to test whether macro-level echo-chamber metrics are more sensitive to the underlying network topology or to seemingly small choices in how agents are activated, exposed to content, and prompted with recent interaction history.

\newpage
\section{Additional Simulation Details}
\label{appendix:simulation-details}

\subsection{Shared Experimental Protocol}

Across both case studies, we use \texttt{gpt-5.2} as the primary model for all reported analyses and replicate the experiments on three additional state-of-the-art, frontier models (\texttt{claude-haiku-4-5}, \texttt{gemini-2.5-flash}, and \texttt{deepseek-v3.2}) as a cross-model robustness check. We prioritize frontier models because they offer the most reliable foundation for creating realistic social simulations. If sensitivity occurs within these advanced models, it indicates a significant finding rather than a simple lack of model performance or intelligence. Results for additional models are provided in Appendix \ref{appendix:extended-results}.

Each condition is repeated with $N=30$ independent simulation runs using distinct random seeds, and the simulation run is the unit of analysis. We report pairwise mean differences, 95\% confidence intervals, and effect sizes measured by Cohen's $d$. Following conventional benchmarks, we interpret $d \approx 0.2$ as small, $d \approx 0.5$ as medium, and $d \approx 0.8$ as large, while treating these values as descriptive guidelines rather than hard cutoffs \citep{cohen2013statistical, funder2019evaluating}. For statistical testing, we use two-sided Mann--Whitney U tests instead of pairwise $t$-tests because several outcomes are bounded, discrete, or have zero run-level variance in some conditions, making variance-based tests undefined or unreliable. Reported $p$-values are unadjusted for multiple comparisons.

Across both case studies, we hold decoding and parsing settings fixed within each comparison so that observed differences can be attributed to the intended perturbation rather than to changes in generation parameters. Unless otherwise specified, model calls use temperature $=0.3$ and top-$p=0.95$. Temperature controls the randomness of token sampling, while top-$p$ controls the cumulative probability mass from which tokens are sampled. In all experiments, agents are required to return a structured JSON response containing their selected action and a brief reason.

\subsection{Prisoner's Dilemma Simulation}

\begin{figure}[H]
    \centering
    \includegraphics[width=0.9\textwidth]
    {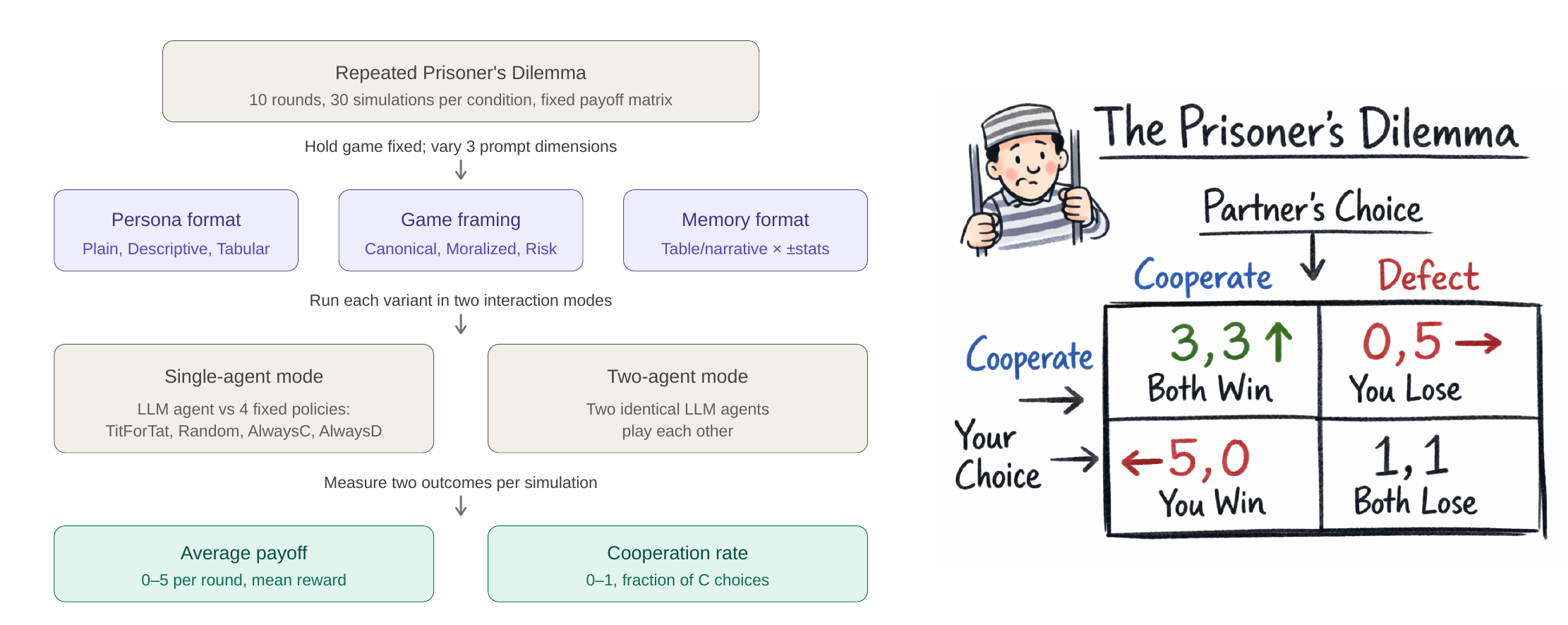}
    \caption{
    \textbf{Repeated Prisoner's Dilemma case-study design.}
    We hold the underlying game fixed across conditions: agents play a 10-round repeated Prisoner's Dilemma with the same payoff matrix, using 30 independent simulations per condition. We then vary three prompt dimensions---persona format, game-instruction framing, and memory format---and run each variant in two interaction modes: a single-agent mode, where one LLM agent plays against four fixed policies, and a two-agent mode, where two identically configured LLM agents play each other. Outcomes are measured using average payoff and cooperation rate.
    }
    \label{fig:prisoner_dilemma_illustration}
\end{figure}

\subsubsection{Setup and Configurations}

Each Prisoner's Dilemma episode runs for $T=10$ rounds. In each round, each player simultaneously chooses Cooperate ($C$) or Defect ($D$). The action space is fixed to \textsc{Cooperate} and \textsc{Defect}, with payoffs $(C,C)=(3,3)$, $(C,D)=(0,5)$, $(D,C)=(5,0)$, and $(D,D)=(1,1)$.

Following common practice in previous studies \citep{akata2025repeatedgames, fontana2025nicer, di2023recognition}, we run two interaction modes. In the \emph{single-agent mode}, one LLM-controlled agent plays against four fixed external policies: \textbf{TitForTat}, which cooperates on the first round and then mirrors the opponent's previous action; \textbf{Random}, which independently selects $C$ or $D$ with equal probability each round; \textbf{AlwaysCooperate}, which unconditionally plays $C$ in every round; and \textbf{AlwaysDefect}, which unconditionally plays $D$ in every round. In the \emph{two-agent mode}, two LLM-controlled agents with identical configurations play one another.

For two-agent runs, both agents are queried once per round, and the prompt order is randomized across rounds to reduce stable positional bias. If the model response cannot be parsed as a valid action after one retry, the simulation falls back to \textsc{Defect}; this conservative fallback avoids dropping failed episodes while preserving a fixed action space. These defaults are implemented consistently across the persona-format, game-instruction, and memory-construction blocks.

\subsubsection{Evaluation Metrics}

For each agent $i$, let $a_{i,t} \in \{C,D\}$ denote the action chosen by agent $i$ in round $t$, where $C$ indicates cooperation and $D$ indicates defection. Let $u_{i,t}$ denote the payoff received by agent $i$ in that round.

\paragraph{Average Payoff.}
Average payoff measures the agent's overall performance under the game's incentive structure:
\[
\text{AveragePayoff}_i
=
\frac{1}{T}
\sum_{t=1}^{T} u_{i,t}.
\]

\paragraph{Cooperation Rate.}
Cooperation rate measures how often the agent chooses cooperation across the repeated game:
\[
\text{CooperationRate}_i
=
\frac{1}{T}
\sum_{t=1}^{T}
\mathbb{I}(a_{i,t}=C).
\]

Here, $T=10$ is the number of rounds and $\mathbb{I}(\cdot)$ is an indicator function equal to 1 when the condition is true and 0 otherwise.

\subsubsection{Perturbation Details}

We test three perturbations, each targeting a design choice that is inconsistently reported and rarely ablated across the LLM-cooperation literature: persona format, game instruction framing, and memory representation. While \citet{fontana2025nicer} carefully document their prompting pipeline and study memory window size in isolation, and \citet{akata2025repeatedgames} report their prompt structure, neither paper systematically varies all three dimensions jointly or tests their interaction effects on cooperative behavior.

\paragraph{Persona format.}
Holding the underlying meaning fixed---the agent is strategic and aims to maximize its payoff---we vary only the presentation across three formats: a paragraph of plain prose (\textsc{Plain}), a descriptive bullet list (\textsc{Descriptive}), and a structured key--value table (\textsc{Tabular}). The corresponding persona texts are reproduced in Figure~\ref{fig:persona_format_panel}.

\paragraph{Game-instruction framing.}
We hold the payoff matrix and persona fixed, using \textsc{plain} as the default persona, and vary only the framing of the game instructions. We test three framings: canonical game-theory framing (\textsc{Canonical}), which uses standard labels such as ``Cooperate'' and ``Defect''; moralized framing (\textsc{Moralized}), which describes the same actions as ``cooperate fairly'' versus ``exploit''; and risk framing (\textsc{Risk}), which presents cooperation as safer but vulnerable and defection as riskier but potentially higher-reward.

\paragraph{Memory representation.}
We hold the persona fixed to \textsc{Plain} and the game framing fixed to \textsc{Canonical}, and vary only how the interaction history is shown to the agent. The design is $2 \times 2$: history format (table vs.\ narrative) crossed with whether summary statistics are included, such as cooperation rates, cumulative payoffs, and joint outcome counts.

\subsection{Echo-Chamber Simulation}

\begin{figure}[H]
    \centering
    \includegraphics[width=0.95\textwidth]
    {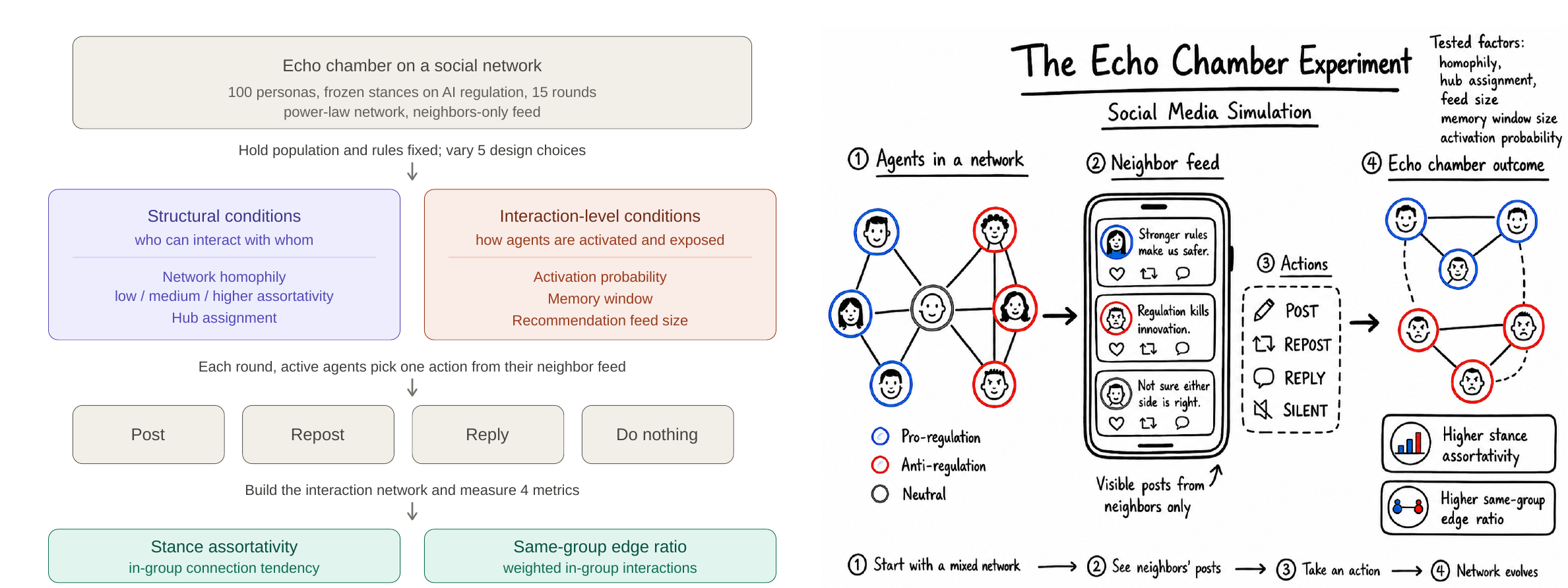}
    \caption{
    \textbf{Echo-chamber case-study design.}
    We simulate 100 LLM agents discussing whether advanced AI systems should be regulated on a fixed social network. Each agent has a persona, follower count, and frozen stance toward AI regulation. In each round, activated agents see recent posts from their direct neighbors only and choose one action: \texttt{POST}, \texttt{REPOST}, \texttt{REPLY}, or \texttt{SILENT}. We vary five architectural perturbations---input-network homophily, hub assignment, recommendation feed size, memory window size, and activation probability---and measure whether these choices shift macro-level echo-chamber outcomes, including stance assortativity and weighted same-group edge ratio.
    }
    \label{fig:echo_chamber_illustration}
\end{figure}

\subsubsection{Setup and Configurations}

The echo-chamber simulation uses 100 agents loaded from a persona file and a fixed edgelist. Each persona includes a short bio and a stance on a focal topic: whether advanced AI systems should be regulated. Stances are represented on a five-point scale, from strongly against regulation to strongly support regulation. Agent stances are frozen throughout the run, so any outcome differences reflect exposure, interaction, and network-structure effects rather than belief updating.

Each run lasts 15 rounds by default. In each round, agents become active independently with activation probability $0.3$; if no agent is sampled as active, one agent is selected at random so that every round contains at least one decision. Each active agent sees recent posts from direct network neighbors only, with a default feed size of 5 and memory window of 2 rounds. The action space is fixed to \textsc{Post}, \textsc{Repost}, \textsc{Reply}, and \textsc{Do Nothing}. Reposts and replies must target a visible feed item; invalid targets are converted to a post or silence depending on feed availability. The default implementation uses 4 worker threads, temperature $=0.3$, and top-$p=0.95$.

\subsubsection{Evaluation Metrics}

Let $G_I=(V,E_I)$ denote the interaction graph induced by observed interactions during a simulation run, where an edge $(i,j)\in E_I$ indicates that agent $i$ reposted or replied to agent $j$. Let $g_i$ denote agent $i$'s stance group, derived from its position on AI regulation: anti-regulation, neutral, or pro-regulation. Let $A_{ij}$ be the weighted adjacency matrix of $G_I$, where $A_{ij}$ is the number of times agent $i$ interacted with agent $j$.

\paragraph{Stance Assortativity.} 
Stance assortativity measures whether agents preferentially interact with agents who hold similar positions on AI regulation. We compute the categorical assortativity coefficient over stance-group labels:
\[
r =
\frac{
\sum_{u \in \mathcal{G}} p_{uu} -
\sum_{u \in \mathcal{G}} p^{\text{out}}_u p^{\text{in}}_u
}{
1 -
\sum_{u \in \mathcal{G}} p^{\text{out}}_u p^{\text{in}}_u
},
\]
where $\mathcal{G}=\{\text{anti-regulation}, \text{neutral}, \text{pro-regulation}\}$ is the set of stance groups. The term $p_{uv}$ denotes the fraction of weighted interaction edges that go from agents in stance group $u$ to agents in stance group $v$:
\[
p_{uv}
=
\frac{
\sum_{i,j} A_{ij}\,\mathbf{1}[g_i=u,\, g_j=v]
}{
\sum_{i,j} A_{ij}
}.
\]
The marginal terms are defined as
\[
p^{\text{out}}_u = \sum_{v \in \mathcal{G}} p_{uv},
\qquad
p^{\text{in}}_u = \sum_{v \in \mathcal{G}} p_{vu}.
\]
Intuitively, $\sum_{u \in \mathcal{G}} p_{uu}$ is the observed fraction of interactions that occur within the same stance group, while $\sum_{u \in \mathcal{G}} p^{\text{out}}_u p^{\text{in}}_u$ is the expected within-group fraction under random mixing that preserves the overall amount of outgoing and incoming interaction for each stance group. The coefficient ranges from $-1$ to $1$: positive values indicate more within-stance-group interaction than expected by chance, values near $0$ indicate approximately random mixing with respect to AI-regulation stance, and negative values indicate more cross-stance-group interaction than expected.

\paragraph{Same-Group Edge Ratio.}
Same-group edge ratio is the fraction of interaction edges that occur between agents with the same side label:
\[
\text{SameGroupRatio}
=
\frac{
\sum_{(i,j)\in E_I} \mathbb{I}(s_i=s_j)
}{
|E_I|
}.
\]

For weighted interaction graphs, we use the weighted version:
\[
\text{SameGroupRatio}_{w}
=
\frac{
\sum_{i,j} A_{ij}\mathbb{I}(s_i=s_j)
}{
\sum_{i,j} A_{ij}
}.
\]

This metric ranges from $0$ to $1$, where $0$ means all interactions are cross-side and $1$ means all interactions are within-side. In a binary-side setting with balanced groups, a value near $0.5$ corresponds to roughly equal within-side and cross-side interaction; values above $0.5$ indicate stronger same-group interaction, while values below $0.5$ indicate more cross-group interaction.

\subsubsection{Perturbation Details}

Prior work highlights network structure, homophily, recommendation-mediated exposure, and memory/context as central mechanisms in echo-chamber formation, making these perturbations natural targets for robustness testing \citep{gu2025echochamber, wang2025decoding, ferraro2024agent}. We test five perturbations that target common but often under-audited design choices in LLM echo-chamber simulations: input-network homophily (\textsc{Low}, \textsc{Medium}, \textsc{High}), hub assignment (\textsc{Anti Hub}, \textsc{Pro Hub}, \textsc{Mixed Hub}, \textsc{Random Hub}), activation probability ($0.3$, $0.5$), memory window ($2$ and $4$ rounds), and recommendation feed size ($5$, $10$ posts). These choices span structural conditions that shape who can interact with whom and interaction-level conditions that shape who becomes active, what content agents see, and what recent context conditions their actions.

\paragraph{How do we decide the fixed degree sequence?}
To hold overall connectivity constant across conditions, we construct a fixed power-law degree sequence for the same set of agents. Specifically, we sample a power-law-like sequence with target mean degree $8$ and exponent $2.4$, then rescale and adjust it until it is graphical. This produces a heavy-tailed network structure in which a small number of agents occupy high-degree positions while most agents have fewer connections. Because the same degree sequence is reused across network variants, differences between conditions cannot be attributed to changes in average degree, density, or the overall presence of hubs.

\paragraph{How do we generate different homophily networks?}
We vary the level of stance homophily while preserving the fixed degree sequence. Let $G_0=(V,E_0)$ denote the input network and let $x_i$ denote agent $i$'s numeric AI-regulation stance on the original five-point scale. We define input-network homophily as the numeric stance assortativity of the input graph:
\[
h(G_0)=
\frac{
\sum_{i,j \in V} A^{(0)}_{ij}(x_i-\bar{x})(x_j-\bar{x})
}{
\sum_{i,j \in V} A^{(0)}_{ij}(x_i-\bar{x})^2
},
\]
where $A^{(0)}_{ij}$ is the adjacency matrix of the input network $G_0$, and $\bar{x}$ is the mean stance among connected endpoints. Intuitively, $h(G_0)$ is high when connected agents tend to have similar stance values and low when edges frequently connect agents with different stances.

Starting from a simple graph that realizes the fixed degree sequence, we apply degree-preserving double-edge swaps to move the graph into a target stance-assortativity band. The rewiring heuristic uses edge-level stance similarity,
\[
\mathrm{sim}(i,j)=1-\frac{|x_i-x_j|}{4},
\]
so that agents with identical stances have similarity $1$, while agents at opposite ends of the five-point scale have similarity $0$. We generate three homophily conditions:
\[
\begin{aligned}
\text{lower homophily} &: \quad h(G_0) \in [0.02, 0.10], \\
\text{medium homophily} &: \quad h(G_0) \in [0.12, 0.20], \\
\text{higher homophily} &: \quad h(G_0) \in [0.22, 0.30].
\end{aligned}
\]
Thus, the homophily manipulation changes which stance groups are connected to one another while preserving each agent's degree.

\paragraph{How do we reassign hubs?}
In a separate manipulation, we vary which agents occupy high-degree hub positions while keeping the same power-law degree sequence. We sort or shuffle agents before assigning degrees, so that the largest degrees can be assigned to different types of agents. In the \textsc{Random} condition, degree assignments are randomly shuffled across agents. In the \textsc{Mixed Hubs} condition, high-degree positions are interleaved across anti-regulation, neutral, and pro-regulation agents, using follower count to rank agents within each stance group. In the \textsc{Pro Hubs} and \textsc{Anti Hubs} conditions, high-degree positions are preferentially assigned to pro-regulation or anti-regulation agents, respectively. After assigning degrees, we again generate a simple graph and rewire it into the same medium homophily band, $h(G_0)\in[0.12,0.20]$. This hub-assignment manipulation changes which agents are structurally central while holding fixed the degree sequence and approximate stance homophily.

\subsection{Recorded Outputs}

We record outputs at both the decision level and the run level. For the Prisoner's Dilemma experiments, each row records one agent decision, including the perturbation condition, seed, round, agent and opponent actions, payoffs, parse-validity indicator, parse note, raw model output, and full prompt input. These fields allow us to compute cooperation rates, payoff trajectories, invalid-output rates, and condition-level behavioral shifts.

For the echo-chamber experiments, we save agent-round logs, round summaries, run metadata, and final agent states. The agent-round logs record each activated agent's profile, network position, action, target post or author when applicable, generated message, exposure diversity, parse validity, full prompt input, and raw model output. Round summaries capture aggregate metrics such as active agents, posts created, cross-group interaction, exposure diversity, graph-side assortativity, polarization, and hub attention. Run metadata records the simulation configuration and input-network properties, while final agent states store each agent's initial and final stance, degree, follower count, and hub status. Together, these outputs support behavioral analysis, robustness diagnostics, and checks of parsing reliability and network-level dynamics.

\newpage
\section{Extended Experimental Results}
\label{appendix:extended-results}

\subsection{Prisoner's Dilemma: Additional Figures}

\begin{figure}[H]
    \centering
    \includegraphics[width=\textwidth]
    {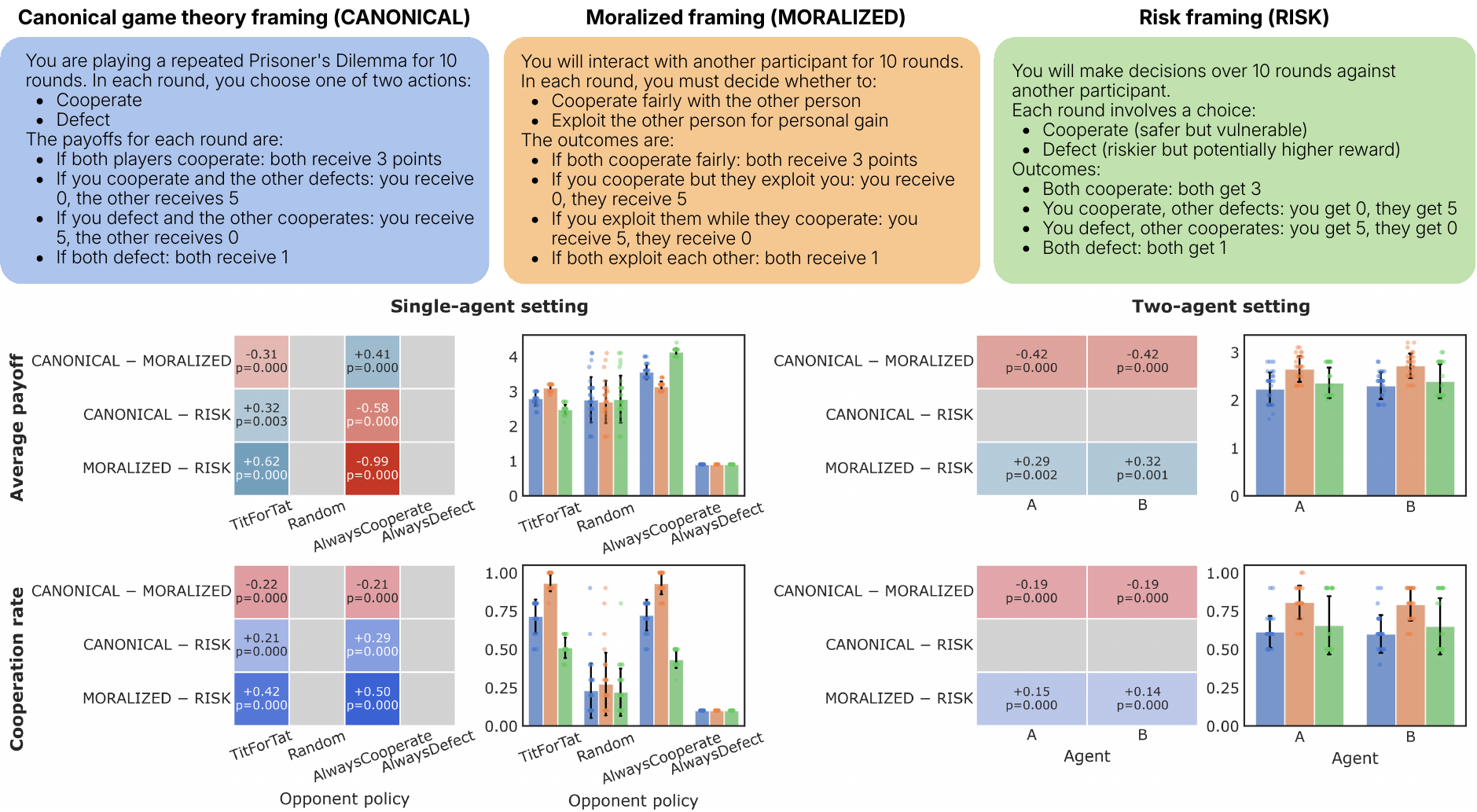}
    \caption{
    \textbf{Effect of game-instruction framing on Prisoner's Dilemma outcomes.}
    Results are shown for the single-agent setting (left) and two-agent setting (right). 
    The heatmaps report statistically significant pairwise differences in average payoff and cooperation rate across instruction framings ($p < .05$, two-sided Mann--Whitney U test), while gray cells indicate non-significant comparisons. 
    Bar plots show run-level distributions for each framing. 
    Bar colors denote instruction framing: \textcolor{blue}{\textsc{canonical}} (blue), \textcolor{orange}{\textsc{moralized}} (orange), and \textcolor{green}{\textsc{risk}} (green).
    }
    \label{fig:game_instruction_panel}
\end{figure}

\begin{figure}[H]
    \centering
    \includegraphics[width=\textwidth]{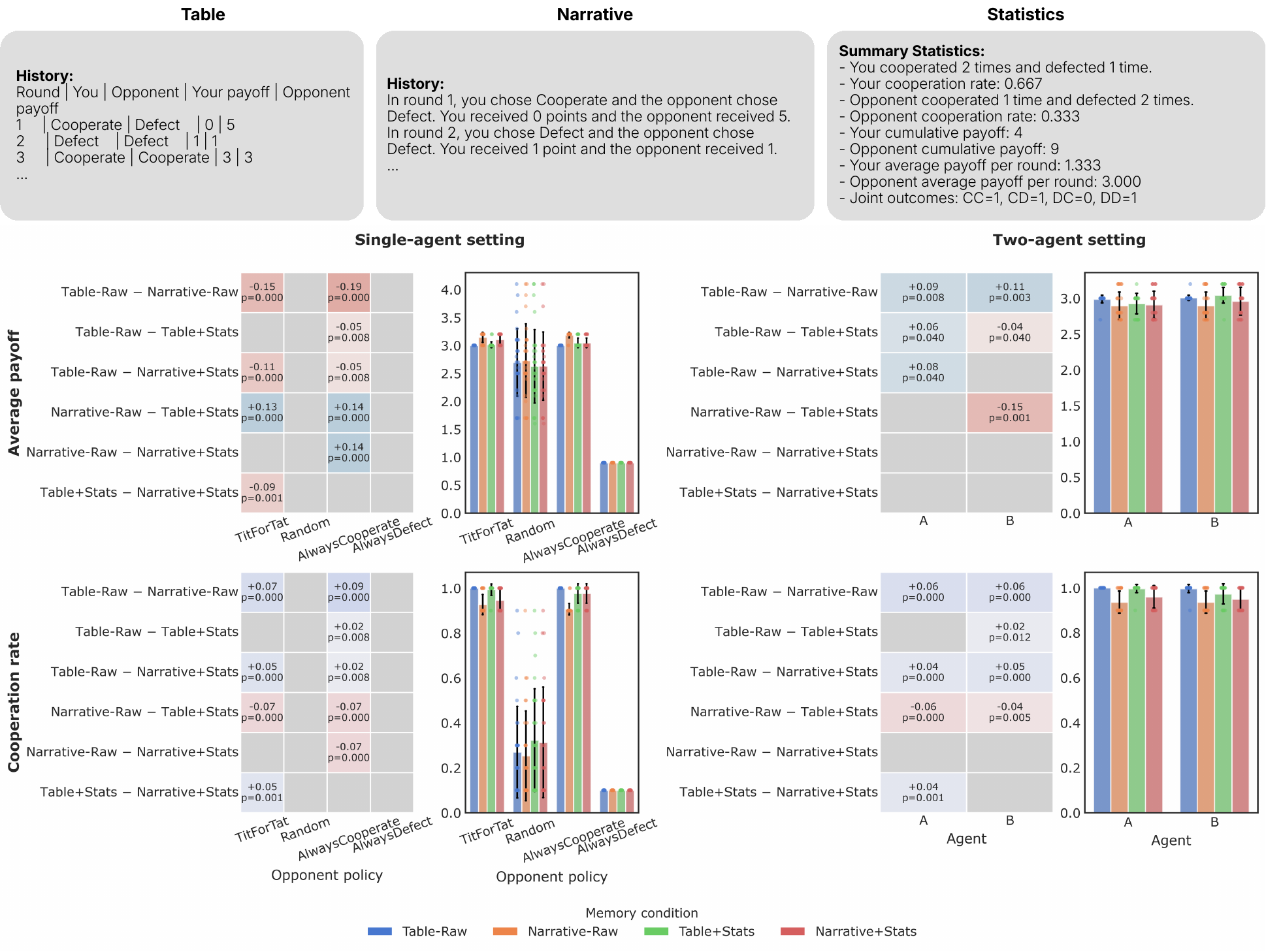}
    \caption{
    \textbf{Effect of memory representation on Prisoner's Dilemma outcomes.}
    Results are shown for the single-agent setting (left) and two-agent setting (right). Memory conditions vary the history format (table vs.\ narrative) and whether summary statistics are included. Heatmaps report statistically significant pairwise differences in average payoff and cooperation rate ($p < .05$, two-sided Mann--Whitney U test), while gray cells indicate non-significant comparisons. Bar plots show run-level distributions for each memory condition.
    }
    \label{fig:memory_representation_panel}
\end{figure}

This section reports the extended Prisoner's Dilemma results for perturbations that are discussed more briefly in the main text. Figure \ref{fig:game_instruction_panel} shows how different framings of the same payoff matrix affect payoff and cooperation, while Figure \ref{fig:memory_representation_panel} shows the effect of changing how prior interaction history is represented. Together, these figures provide the full pairwise comparisons and run-level distributions for the game-instruction and memory-representation perturbations.

\newpage
\vspace*{\fill}
\vspace*{\fill}

\newpage
\subsection{Echo Chamber: Additional Figures}

\begin{figure}[H]
    \centering
    \includegraphics[width=0.85\textwidth]
    {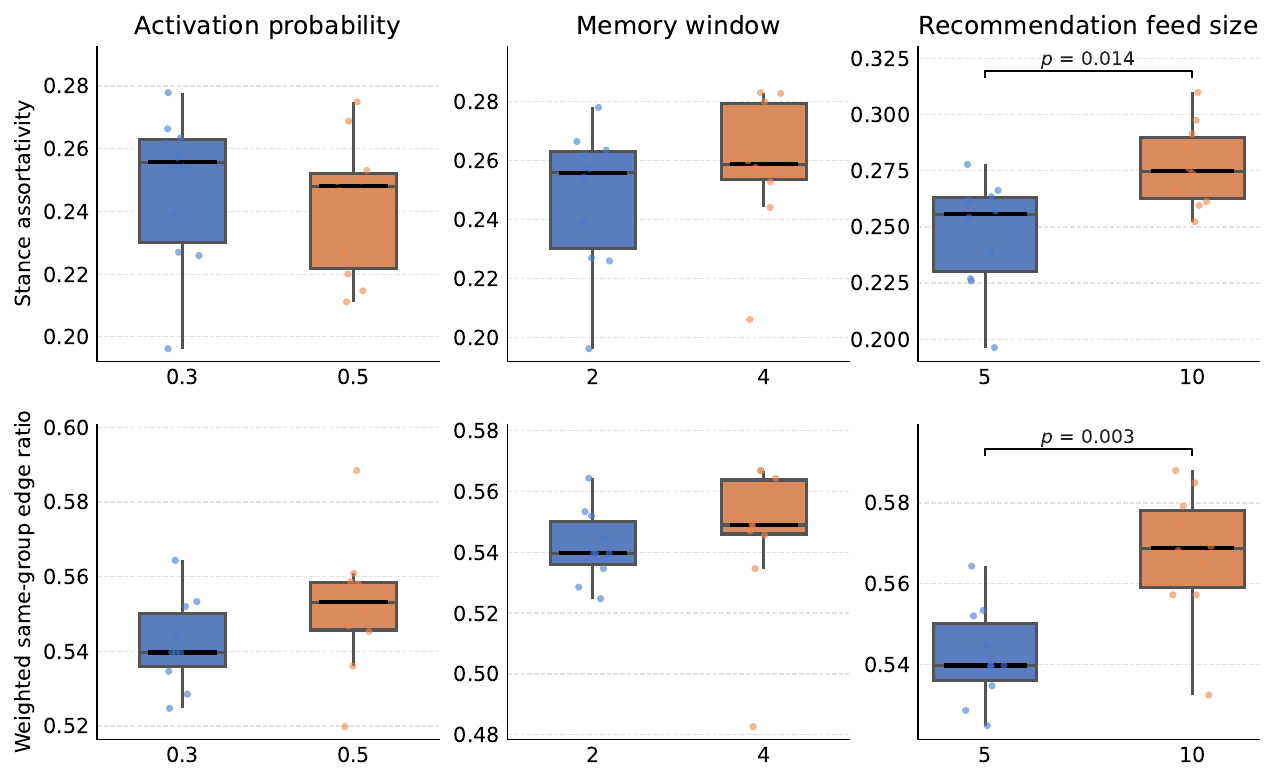}
    \caption{\textbf{Effects of activation probability, memory window, and recommendation feed size on echo-chamber outcomes.}
    Boxplots show the effects of varying activation probability, memory window, and recommendation feed size while holding the input network fixed. The top row reports final stance assortativity and the bottom row reports weighted same-group edge ratio, both computed on the simulated interaction network. Significant pairwise differences are annotated using two-sided Mann--Whitney U tests with $p < 0.05$.}
    \label{fig:prob_window_feed_comparison_gpt}
\end{figure}

\paragraph{Activation probability, memory window, and recommendation feed size.}
Figure~\ref{fig:prob_window_feed_comparison_gpt} reports the effects of varying activation probability, memory window, and recommendation feed size. Activation probability and memory window do not produce statistically significant differences in either final stance assortativity or weighted same-group edge ratio. In contrast, increasing the recommendation feed size from $5$ to $10$ posts significantly increases both stance assortativity ($M=0.247$ to $0.277$, $p=0.014$) and weighted same-group edge ratio ($M=0.542$ to $0.568$, $p=0.003$). This suggests that, among these three parameters, recommendation feed size is the only one that consistently strengthens echo-chamber-like interaction.

\newpage
\subsection{Prisoner's Dilemma: Experiment Results With Other LLMs}

We use \texttt{gpt-5.2} as the primary frontier model in the main text and replicate the Prisoner's Dilemma experiments on three additional state-of-the-art, frontier models: \texttt{claude-haiku-4-5}, \texttt{gemini-2.5-flash}, and \texttt{deepseek-v3.2}. These models are included as a cross-model robustness check. The goal is not to benchmark model quality, but to test whether the observed sensitivity to persona format, game-instruction framing, and memory representation also appears beyond a single model family. If similar sensitivity appears across this diverse set of closed and open-weight model families, then the effect is less likely to be an artifact of one specific model.

\begin{figure}[H]
    \centering
    \includegraphics[width=\textwidth]{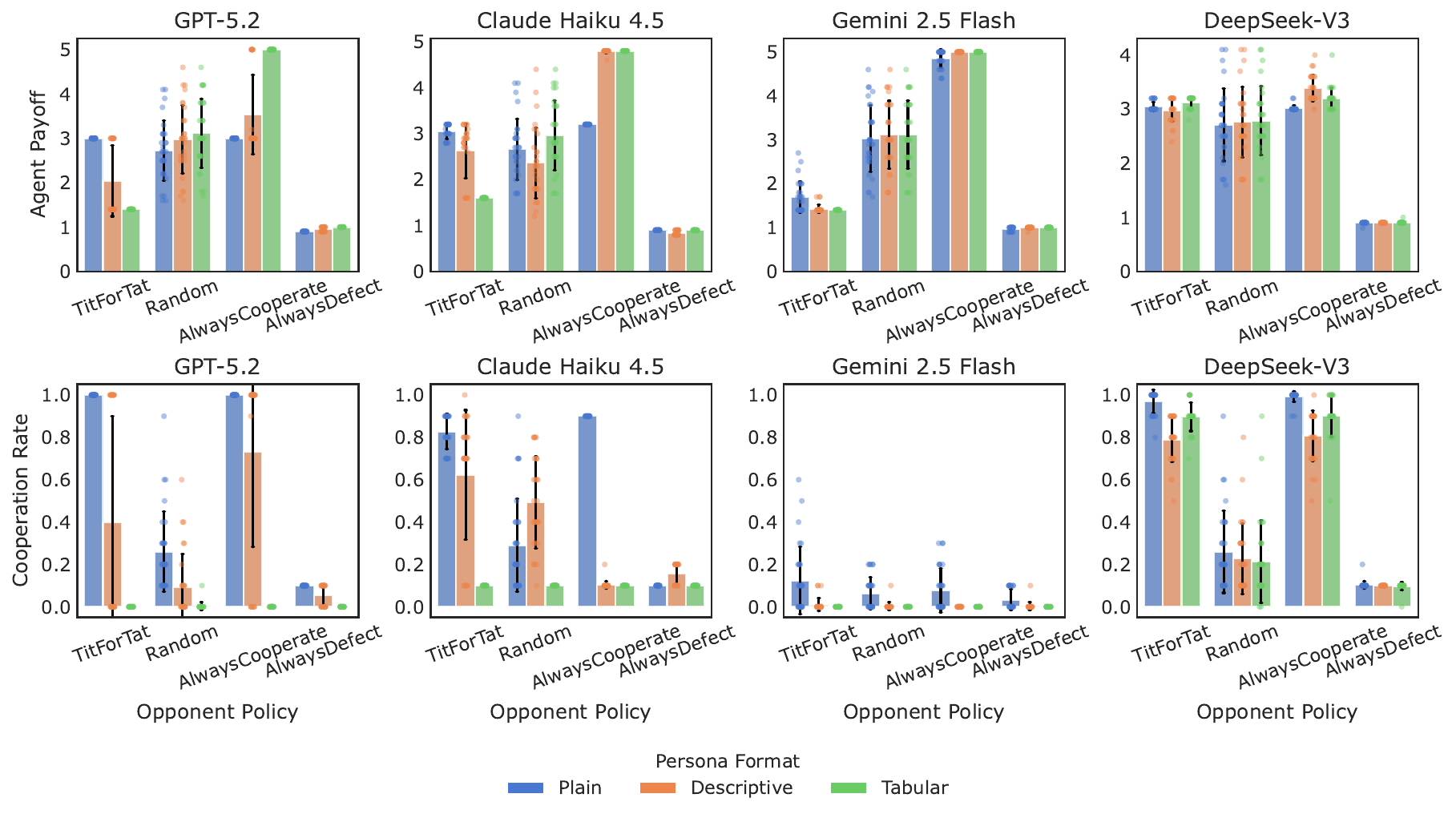}
    \caption{
    \textbf{Persona-format effects in the single-agent Prisoner's Dilemma across four models.}
    Bar plots show average payoff and cooperation rate for each persona format, opponent policy, and model.
    }
    \label{fig:pd_persona_bar}
\end{figure}

\begin{figure}[H]
    \centering
    \includegraphics[width=\textwidth]{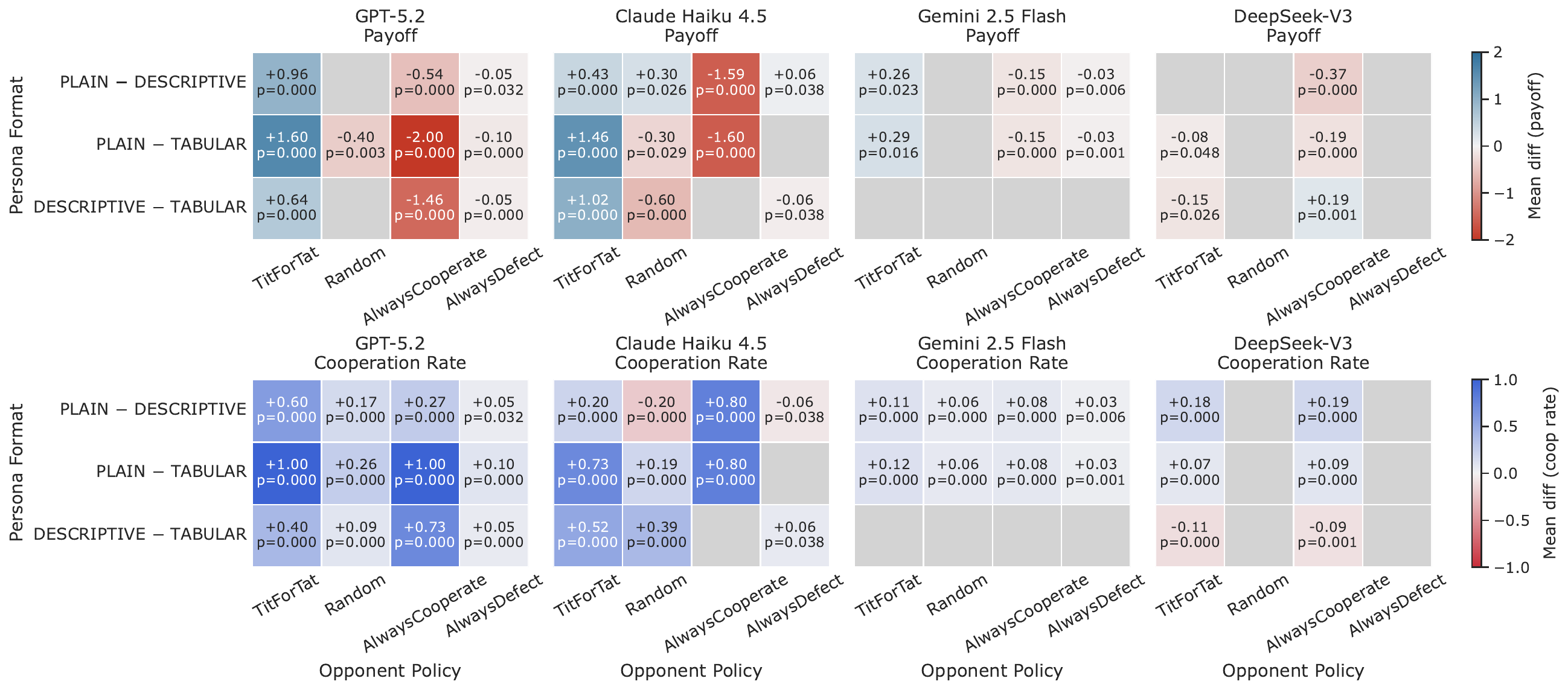}
    \caption{
    \textbf{Pairwise persona-format differences in the single-agent Prisoner's Dilemma across four models.}
    Heatmaps report statistically significant pairwise differences in average payoff and cooperation rate across persona formats.
    }
    \label{fig:pd_persona_heatmap}
\end{figure}

\newpage
\begin{figure}[H]
    \centering
    \includegraphics[width=\textwidth]{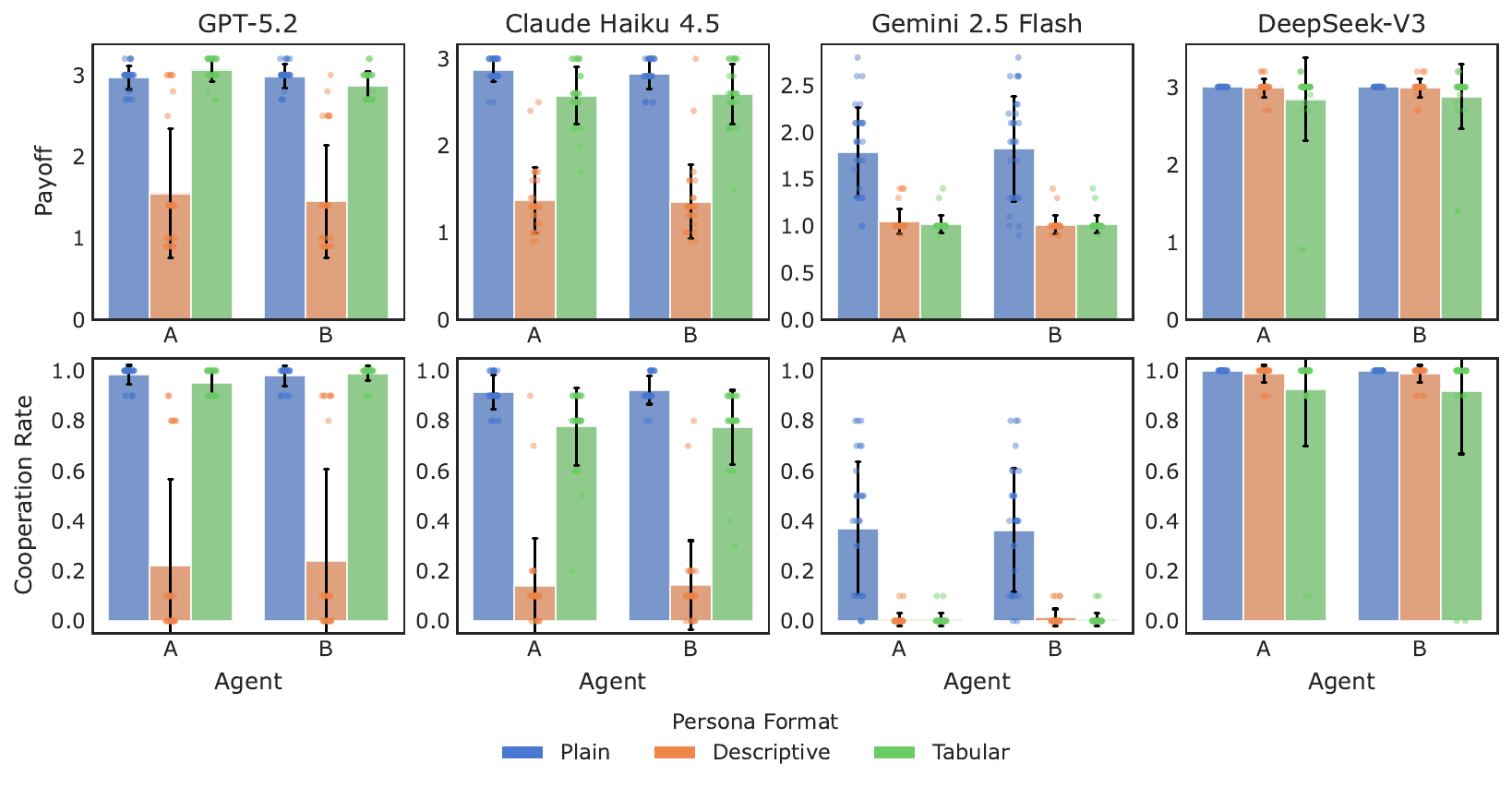}
    \caption{
    \textbf{Persona-format effects in the two-agent Prisoner's Dilemma across four models.}
    Bar plots show average payoff and cooperation rate when both LLM agents use the same persona format.
    }
    \label{fig:pd_persona_bar_two_agent}
\end{figure}

\begin{figure}[H]
    \centering
    \includegraphics[width=\textwidth]{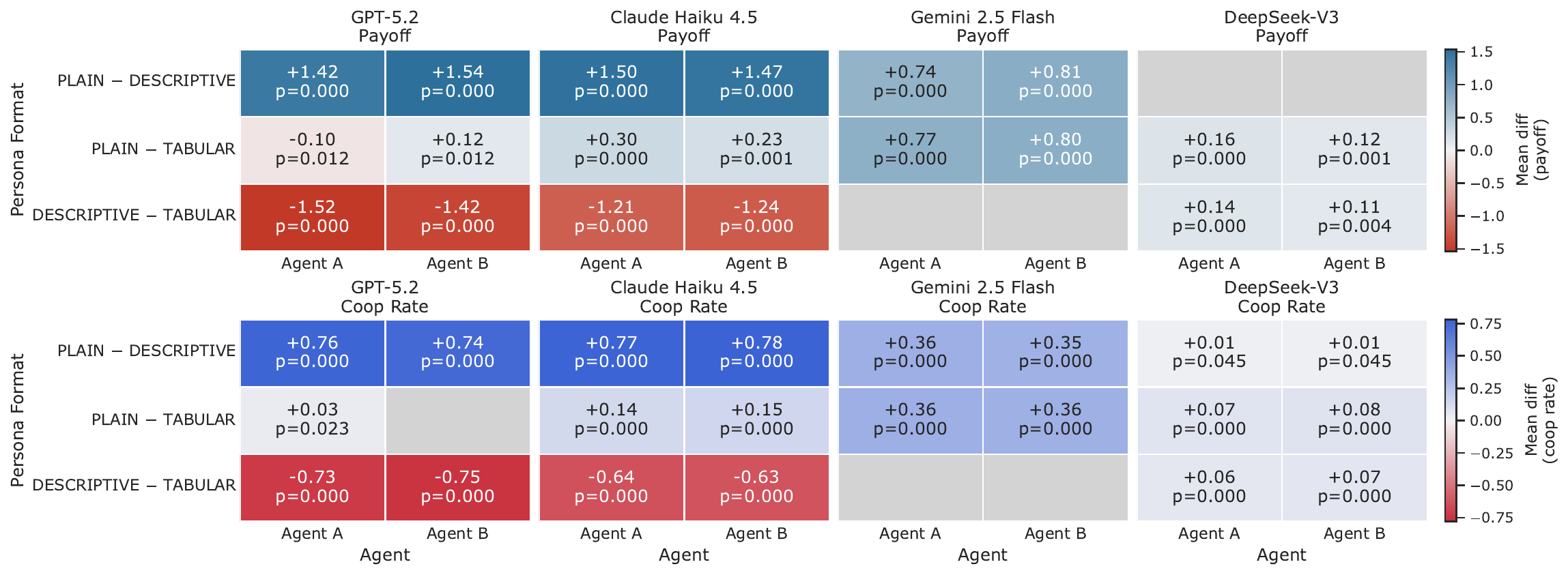}
    \caption{
    \textbf{Pairwise persona-format differences in the two-agent Prisoner's Dilemma across four models.}
    Heatmaps report statistically significant pairwise differences in average payoff and cooperation rate across persona formats.
    }
    \label{fig:pd_persona_heatmap_two_agent}
\end{figure}

\newpage
\begin{figure}[H]
    \centering
    \includegraphics[width=\textwidth]{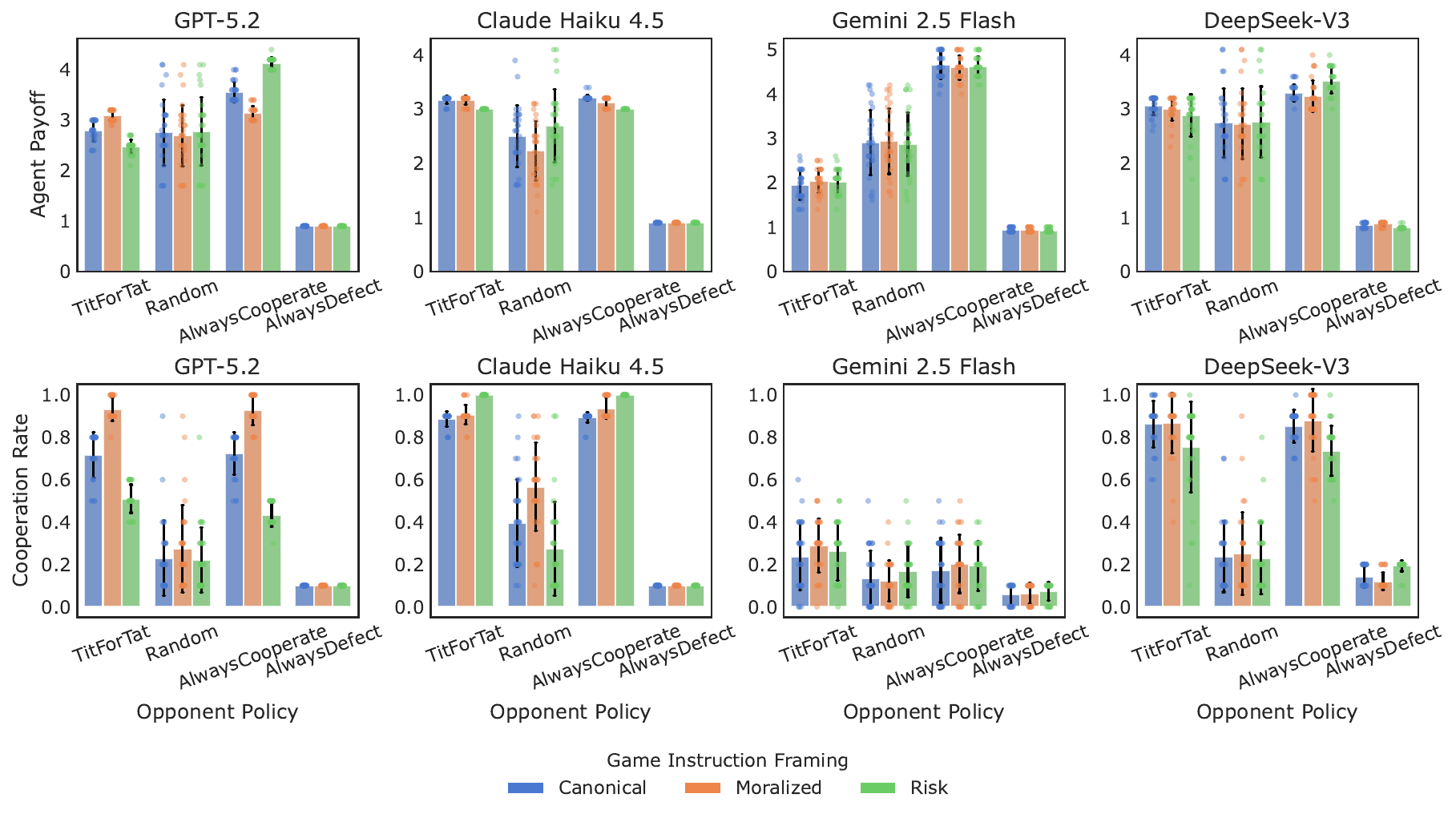}
    \caption{
    \textbf{Game-instruction framing effects in the single-agent Prisoner's Dilemma across four models.}
    Bar plots show average payoff and cooperation rate for each instruction framing, opponent policy, and model.
    }
    \label{fig:pd_framing_bar}
\end{figure}

\begin{figure}[H]
    \centering
    \includegraphics[width=\textwidth]{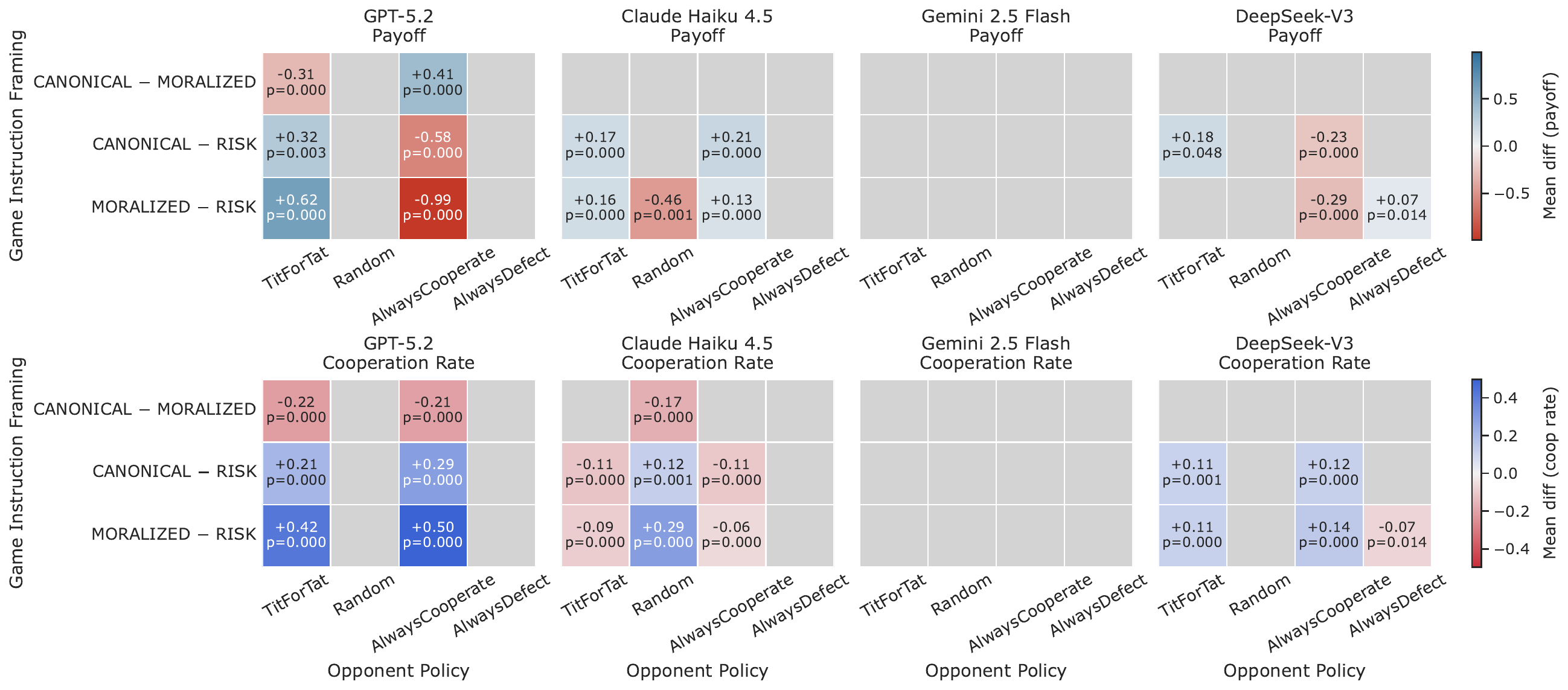}
    \caption{
    \textbf{Pairwise game-instruction framing differences in the single-agent Prisoner's Dilemma across four models.}
    Heatmaps report statistically significant pairwise differences in average payoff and cooperation rate across instruction framings.
    }
    \label{fig:pd_framing_heatmap}
\end{figure}

\newpage
\begin{figure}[H]
    \centering
    \includegraphics[width=\textwidth]{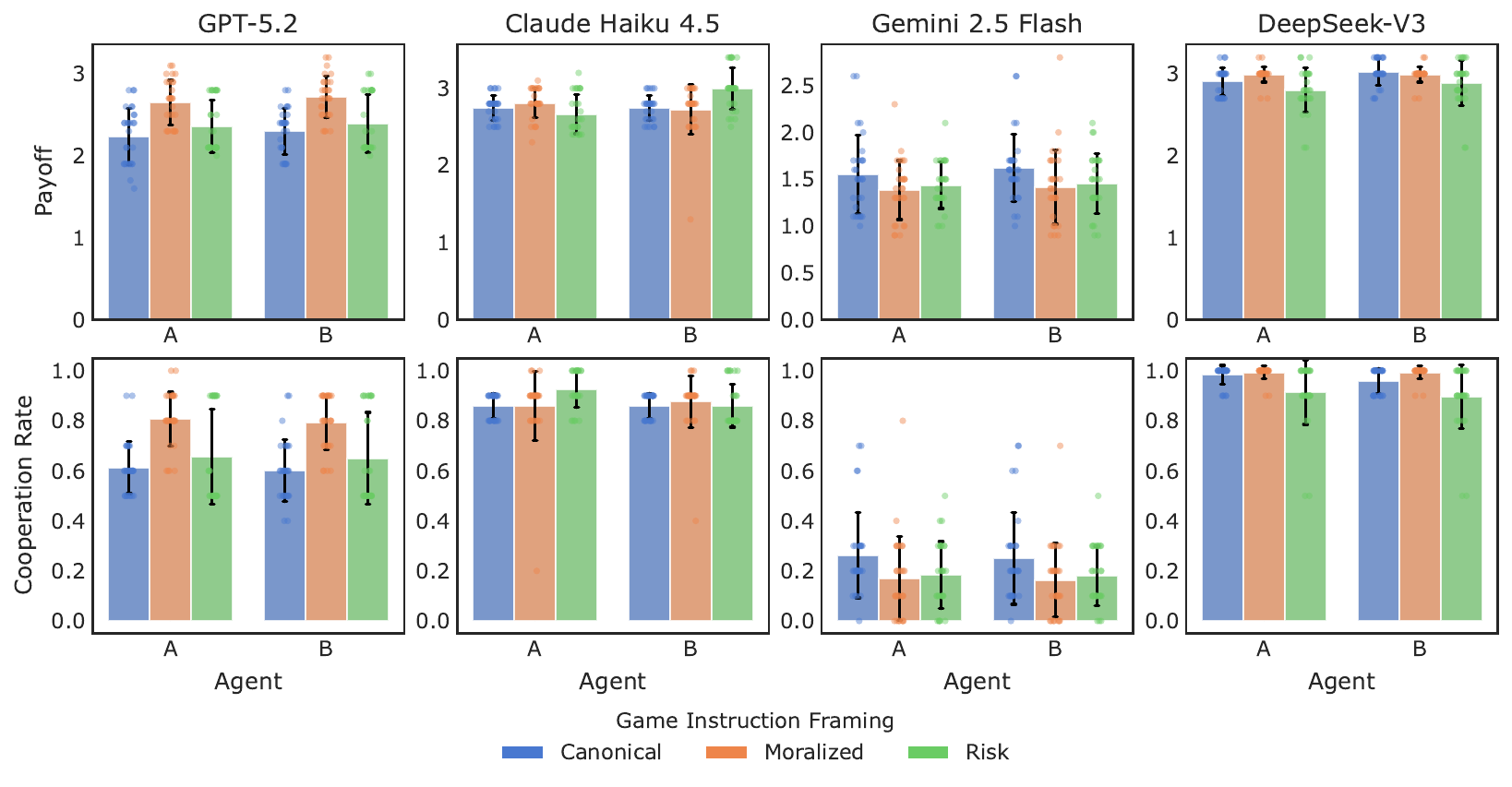}
    \caption{
    \textbf{Game-instruction framing effects in the two-agent Prisoner's Dilemma across four models.}
    Bar plots show average payoff and cooperation rate when both LLM agents receive the same game framing.
    }
    \label{fig:pd_framing_bar_two_agent}
\end{figure}

\begin{figure}[H]
    \centering
    \includegraphics[width=\textwidth]{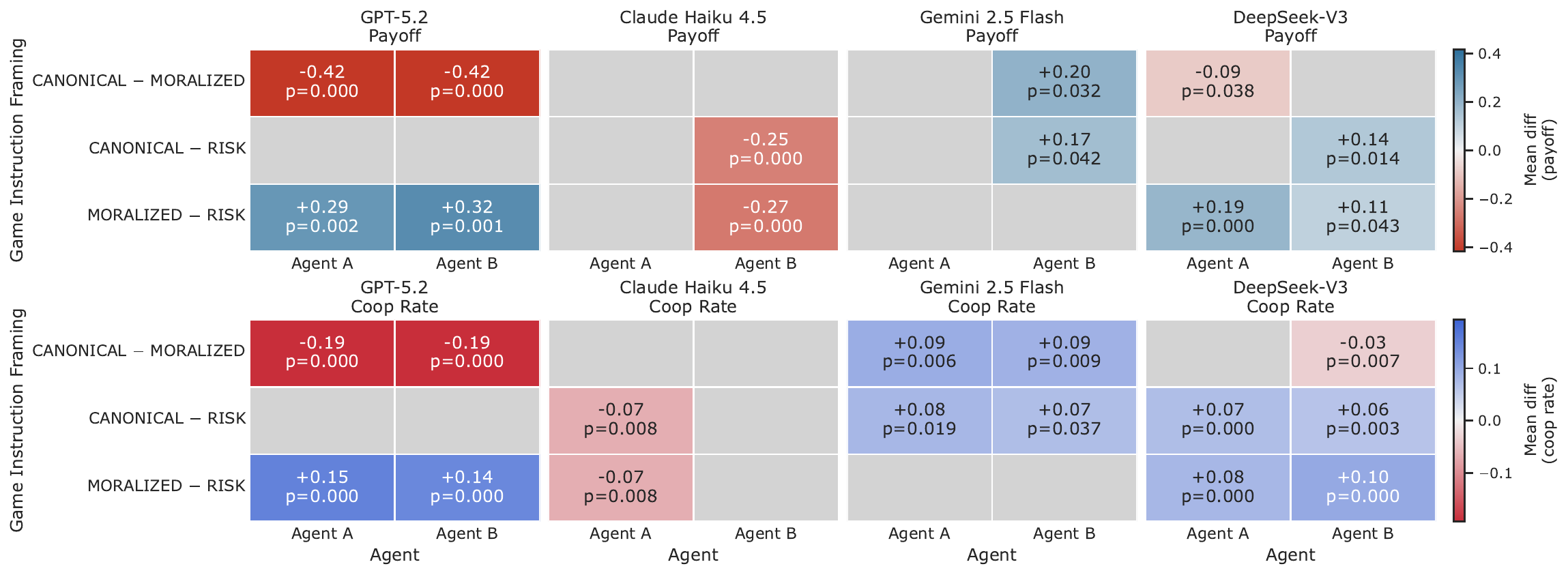}
    \caption{
    \textbf{Pairwise game-instruction framing differences in the two-agent Prisoner's Dilemma across four models.}
    Heatmaps report statistically significant pairwise differences in average payoff and cooperation rate across instruction framings.
    }
    \label{fig:pd_framing_heatmap_two_agent}
\end{figure}

\newpage
\begin{figure}[H]
    \centering
    \includegraphics[width=\textwidth]{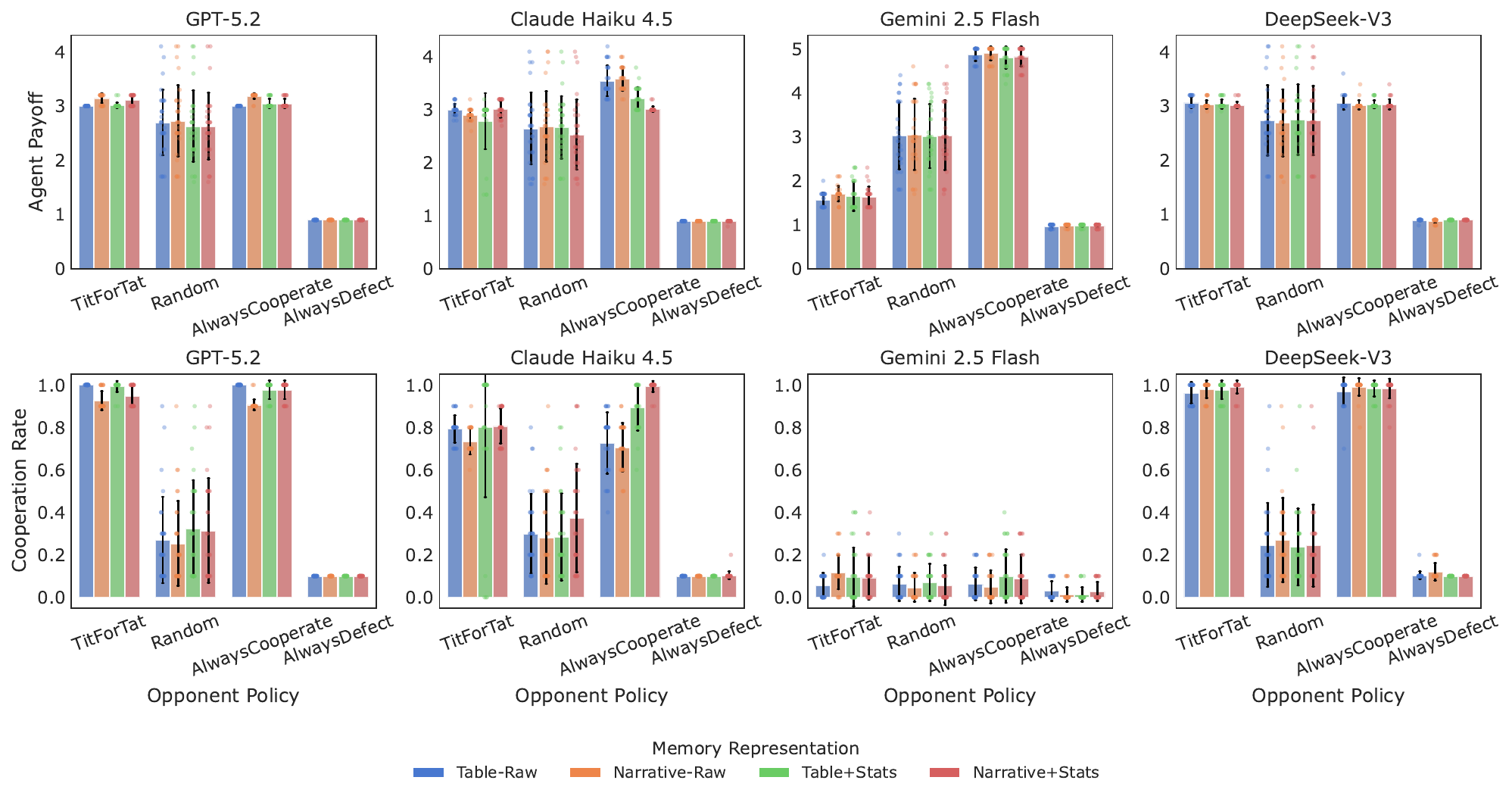}
    \caption{
    \textbf{Memory-representation effects in the single-agent Prisoner's Dilemma across four models.}
    Bar plots show average payoff and cooperation rate for each memory condition, opponent policy, and model.
    }
    \label{fig:pd_memory_bar}
\end{figure}

\begin{figure}[H]
    \centering
    \includegraphics[width=\textwidth]{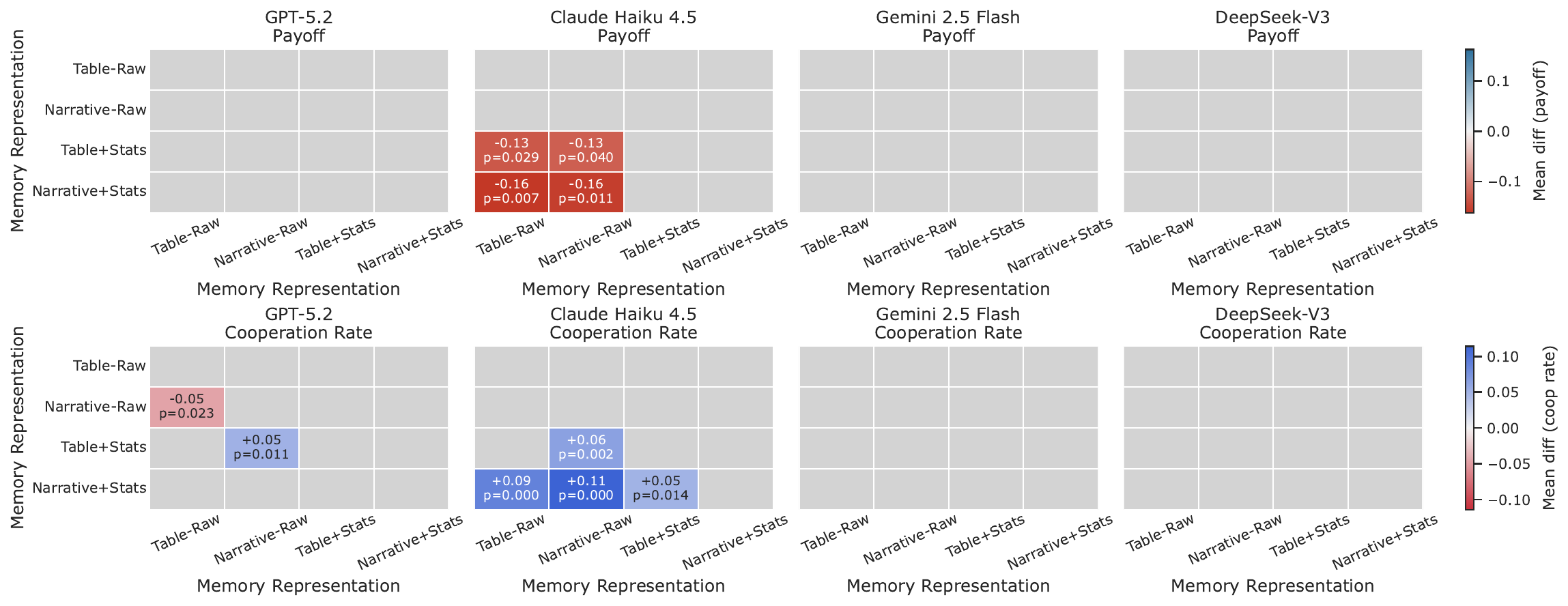}
    \caption{
    \textbf{Pairwise memory-representation differences in the single-agent Prisoner's Dilemma across four models.}
    Heatmaps report statistically significant pairwise differences in average payoff and cooperation rate across memory conditions.
    }
    \label{fig:pd_memory_heatmap}
\end{figure}

\newpage
\begin{figure}[H]
    \centering
    \includegraphics[width=\textwidth]{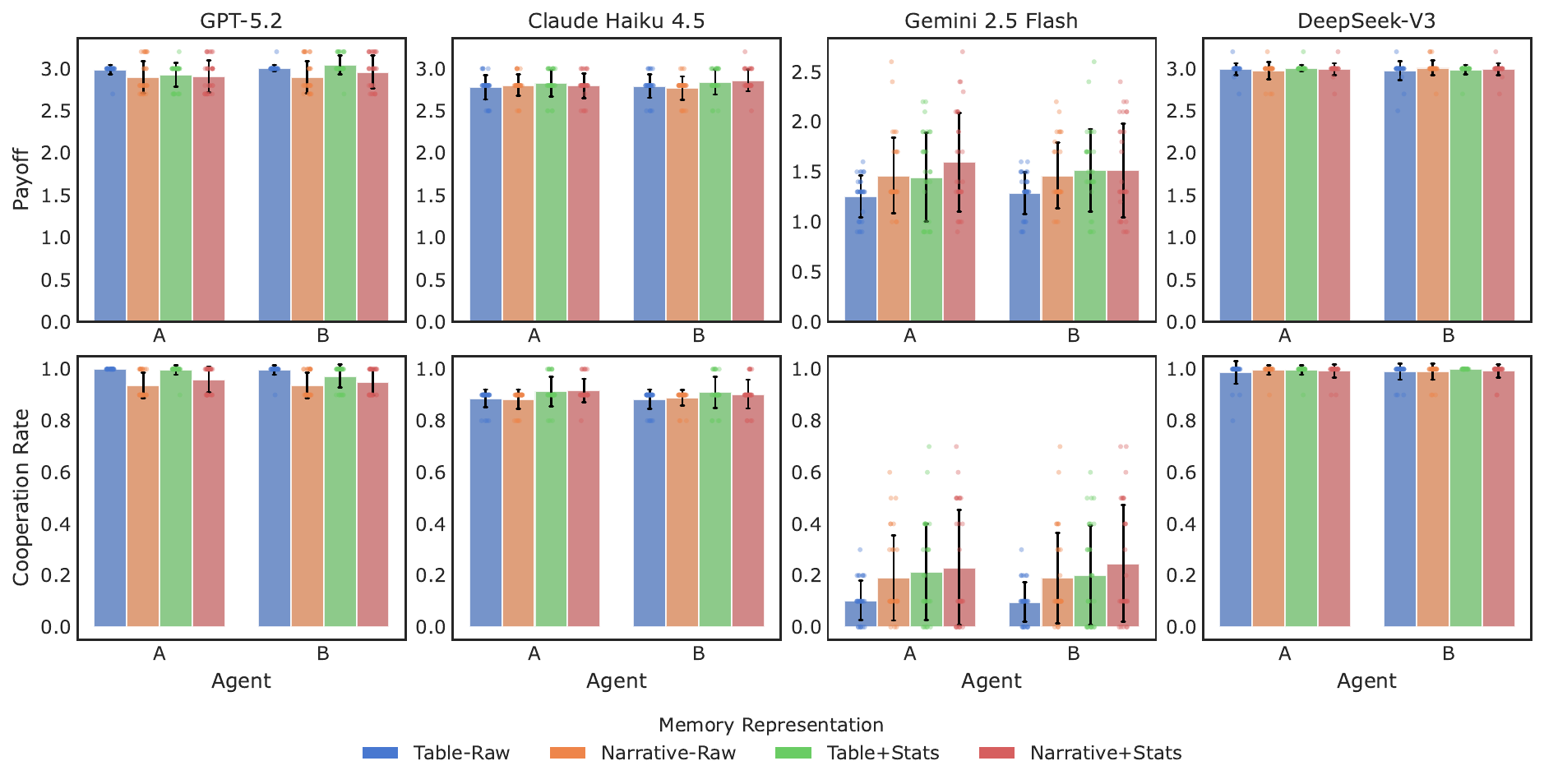}
    \caption{
    \textbf{Memory-representation effects in the two-agent Prisoner's Dilemma across four models.}
    Bar plots show average payoff and cooperation rate when both LLM agents use the same memory representation.
    }
    \label{fig:pd_memory_bar_two_agent}
\end{figure}

\begin{figure}[H]
    \centering
    \includegraphics[width=\textwidth]{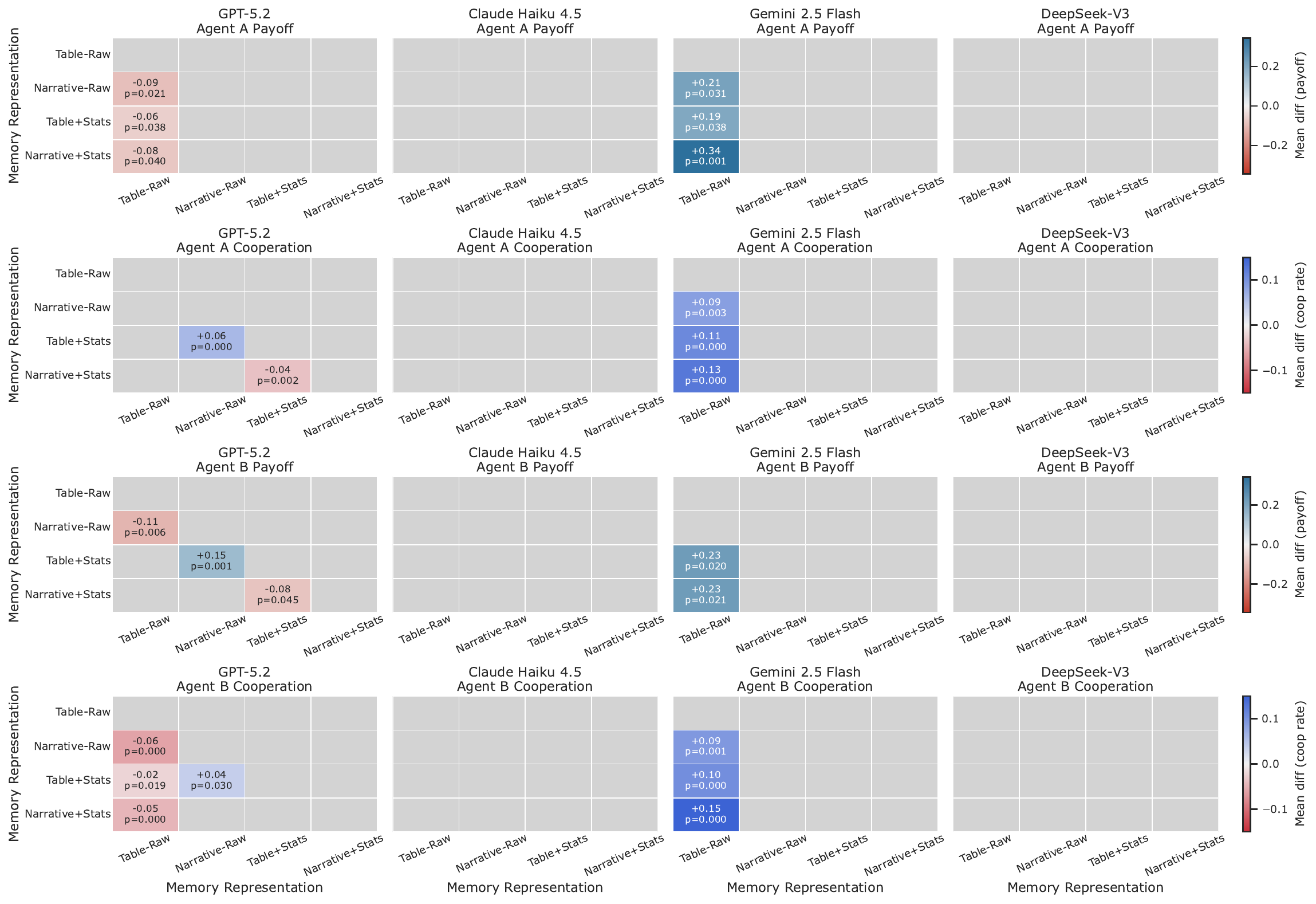}
    \caption{
    \textbf{Pairwise memory-representation differences in the two-agent Prisoner's Dilemma across four models.}
    Heatmaps report statistically significant pairwise differences in average payoff and cooperation rate across memory conditions.
    }
    \label{fig:pd_memory_heatmap_two_agent}
\end{figure}

\newpage
\subsection{Echo Chamber: Experiment Results With Other LLMs}

We additionally replicate the echo chamber experiments with \texttt{claude-haiku-4-5}, \texttt{gemini-2.5-flash}, and \texttt{deepseek-v3.2}. This cross-model check tests whether sensitivity to network structure and exposure design persists beyond the primary \texttt{gpt-5.2} runs, rather than reflecting the behavior of one model family alone.

\begin{figure}[H]
    \centering
    \includegraphics[width=\textwidth]{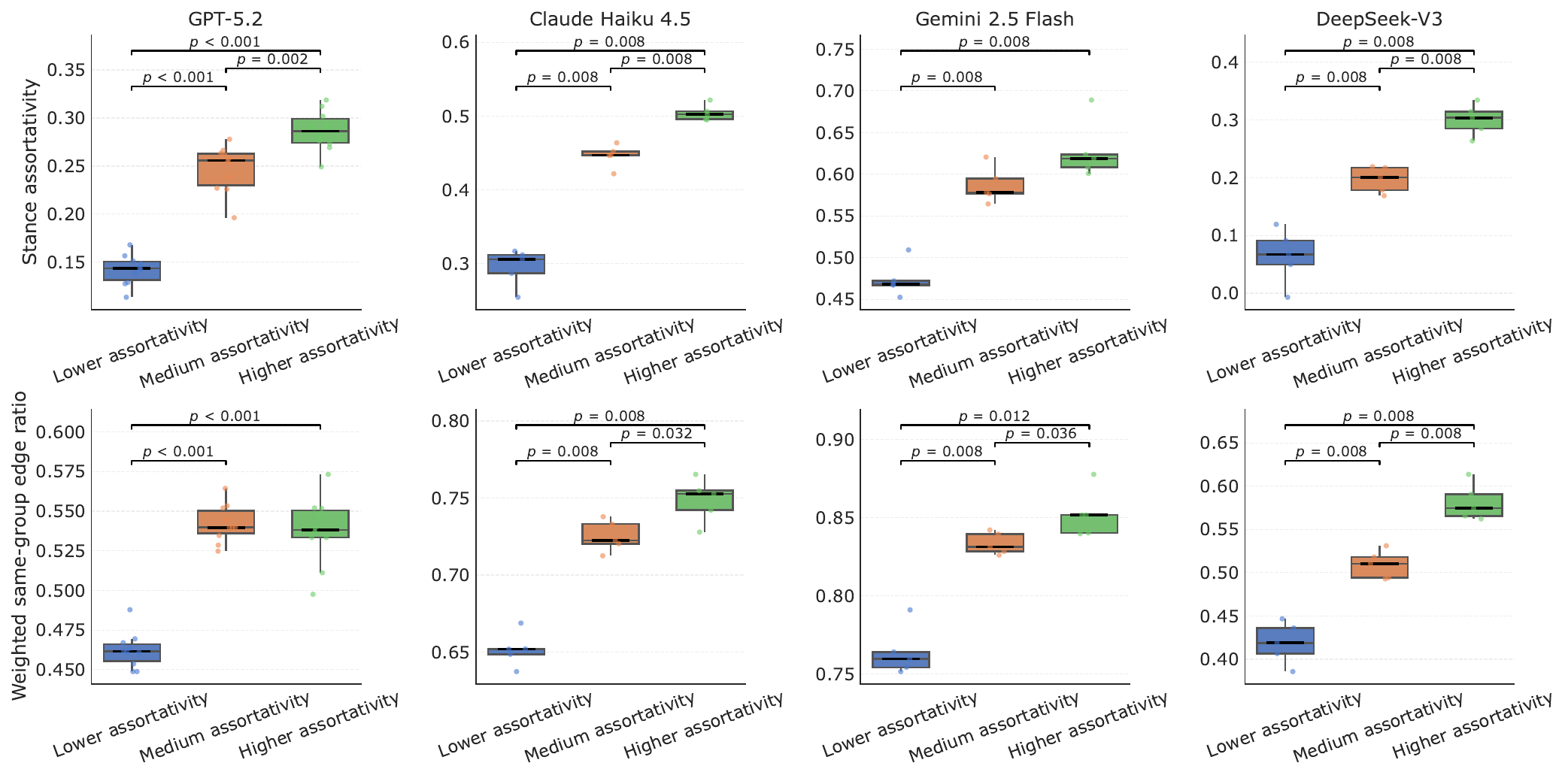}
    \caption{\textbf{Cross-model results for initial network homophily.}
    Boxplots compare final stance assortativity and weighted same-group edge ratio across input networks with different initial stance assortativity levels. Results are shown separately for \texttt{gpt-5.2}, \texttt{claude-haiku-4-5}, \texttt{gemini-2.5-flash}, and \texttt{deepseek-v3.2}.}
    \label{fig:model_comparison_assortativity_metrics}
\end{figure}

\newpage
\begin{figure}[H]
    \centering
    \includegraphics[width=\textwidth]{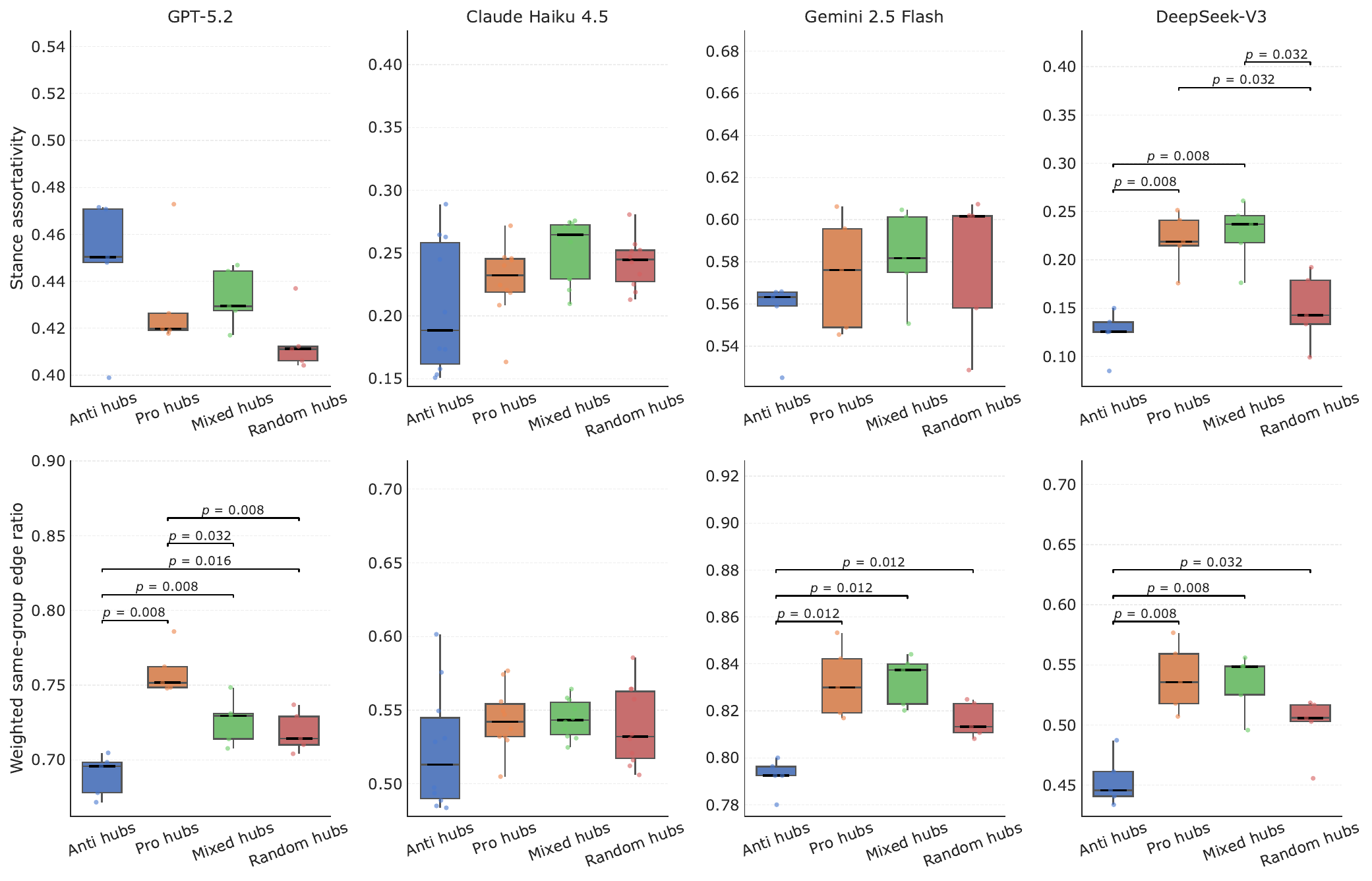}
    \caption{\textbf{Cross-model results for hub assignment.}
    Boxplots compare final stance assortativity and weighted same-group edge ratio when the highest-degree nodes are assigned to anti-regulation, pro-regulation, mixed, or random agents. Results are shown separately for each model.}
    \label{fig:model_comparison_hubs_metrics}
\end{figure}

\begin{figure}[H]
    \centering
    \includegraphics[width=\textwidth]{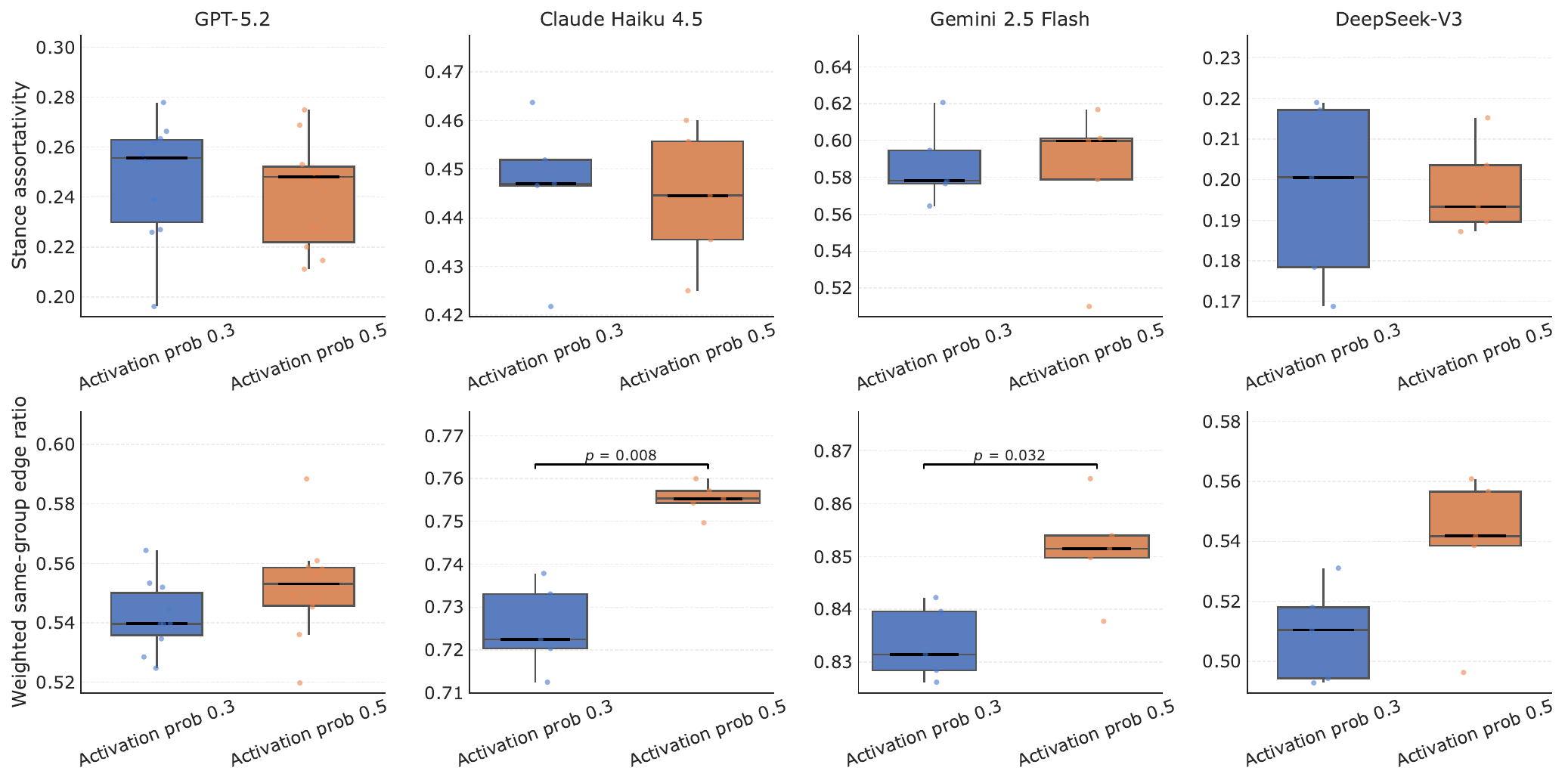}
    \caption{\textbf{Cross-model results for activation probability.}
    Boxplots compare final stance assortativity and weighted same-group edge ratio under different agent activation probabilities. Results are shown separately for each model.}
    \label{fig:model_comparison_activationprob_metrics}
\end{figure}

\newpage
\begin{figure}[H]
    \centering
    \includegraphics[width=\textwidth]{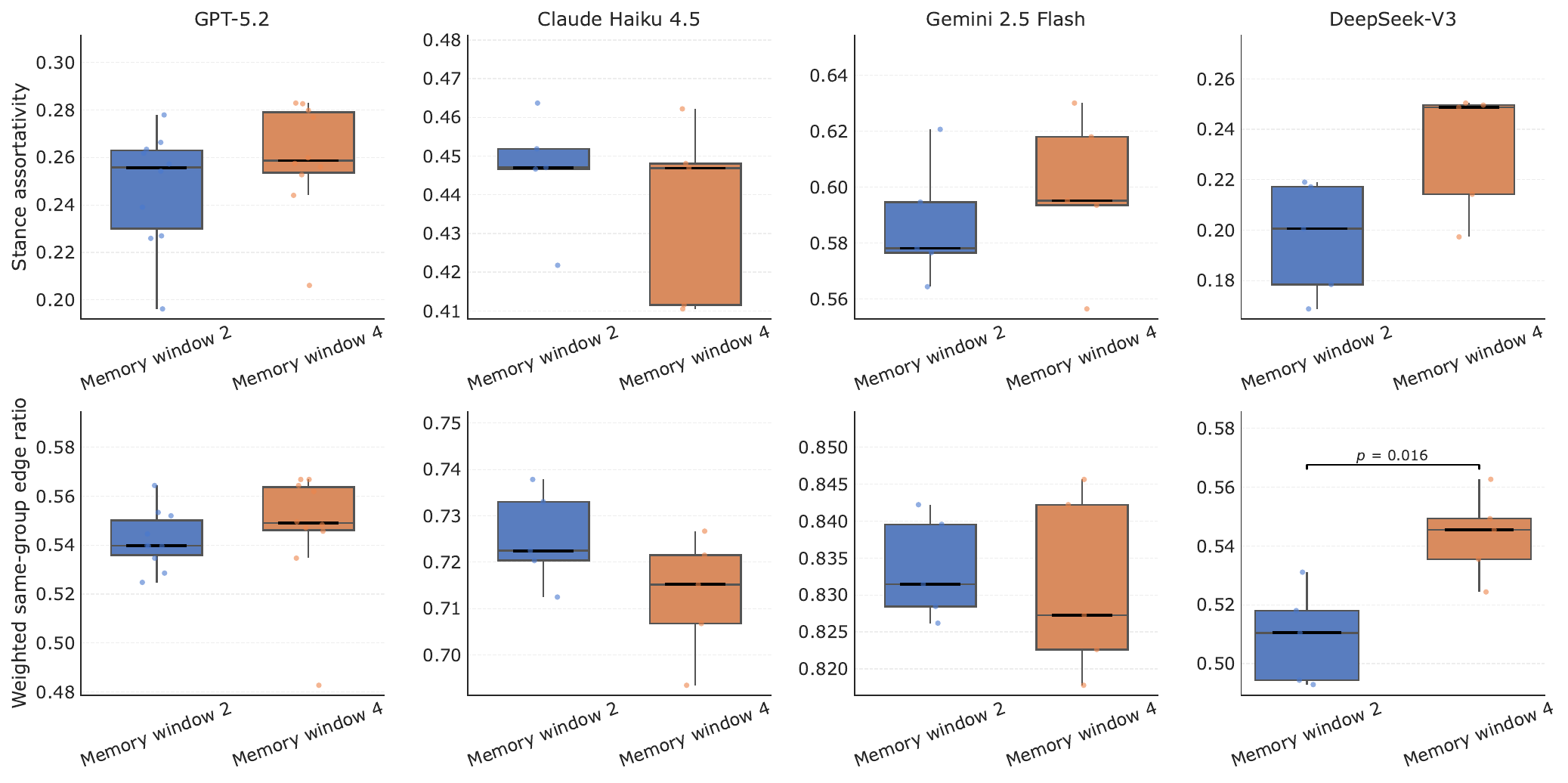}
    \caption{\textbf{Cross-model results for memory window.}
    Boxplots compare final stance assortativity and weighted same-group edge ratio under different memory-window lengths. Results are shown separately for each model.}
    \label{fig:model_comparison_memory_metrics}
\end{figure}

\begin{figure}[H]
    \centering
    \includegraphics[width=\textwidth]{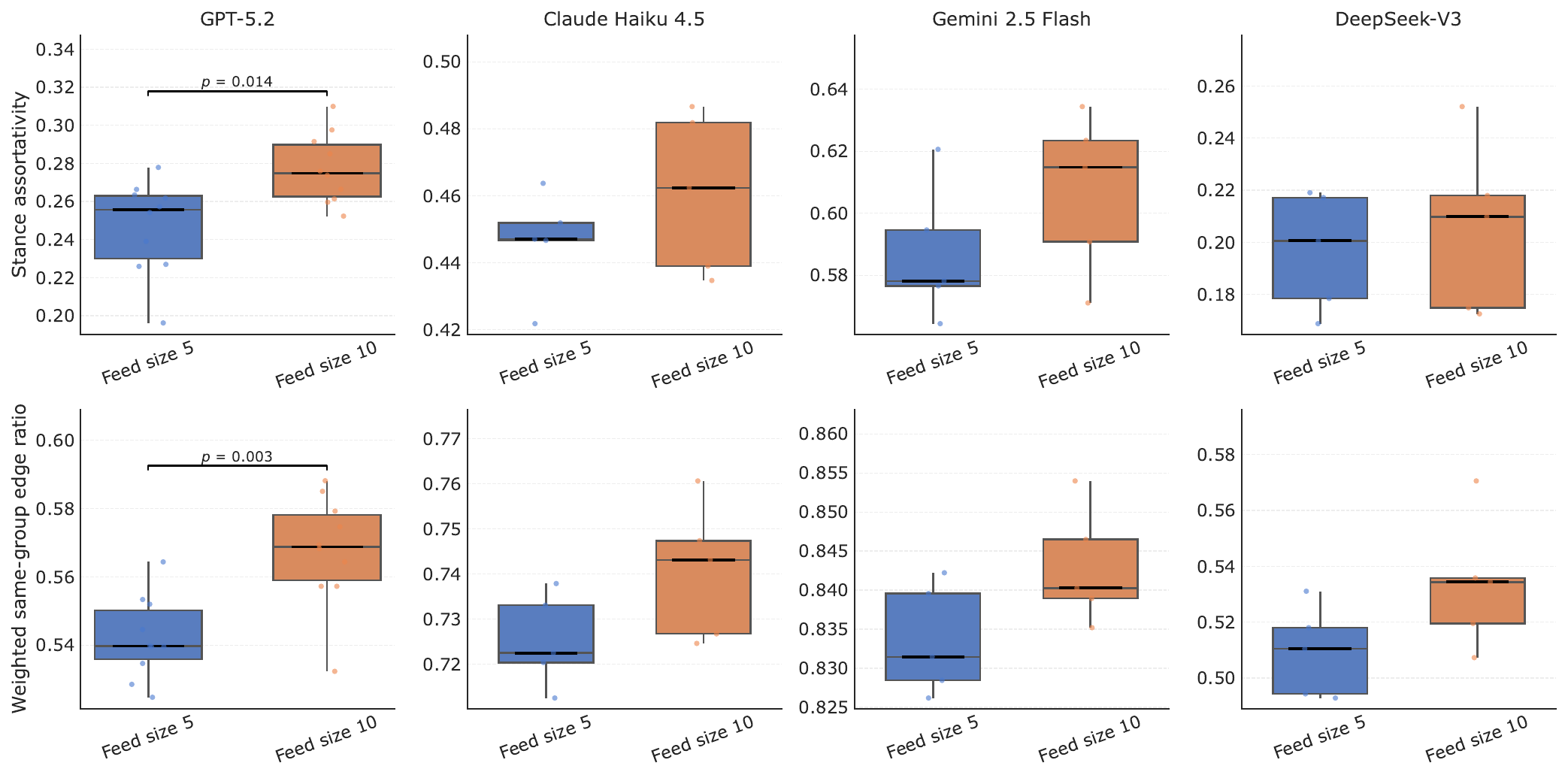}
    \caption{\textbf{Cross-model results for recommendation feed size.}
    Boxplots compare final stance assortativity and weighted same-group edge ratio when agents are exposed to different numbers of recent neighbor posts. Results are shown separately for each model.}
    \label{fig:model_comparison_feedsize_metrics}
\end{figure}

\newpage
\section{The Social Simulation Audit Card}
\label{appendix:audit-card}

The card links TRAILS perturbations to a claim's scope: context-relevant social or theoretical conditions define that scope, whereas context-irrelevant implementation choices should not determine the conclusion.

\begin{enumerate}[leftmargin=*,nosep]
\item \textbf{State the claim:} specify the conclusion, evidentiary role, and property to preserve.
\item \textbf{Map perturbations:} identify TRAILS-D or TRAILS-R and justify context relevance.
\item \textbf{Register the audit:} specify alternatives, controls, models, runs, and preservation criteria before examining results.
\item \textbf{Report sensitivity:} give effect sizes and uncertainty; statistical significance alone does not establish claim preservation.
\item \textbf{Scope the conclusion:} report it as \textit{Robust}, \textit{Scope-limited}, \textit{Implementation-sensitive}, or \textit{Inconclusive}, bounded by the tested conditions and remaining uncertainty.
\end{enumerate}

These are recommended reporting steps, not a claim that the present case studies constitute preregistered or exhaustive audits. Generalization to human behavior additionally requires external validation.

\subsection{Auditing Scientific Claims in the Two Case Studies}
\label{appendix:case-study-audits}

\begin{table*}[t]
\centering
\scriptsize
\setlength{\tabcolsep}{4pt}
\caption{Claim-level interpretations of the perturbations reported in this paper for GPT-5.2. Context relevance is relative to claims about cooperation or structural echo chambers. This qualitative summary follows Section~\ref{sec:case_study}; it does not introduce additional experiments or statistical tests.}
\label{tab:case-study-perturbations}
\begin{tabular}{@{}
    >{\raggedright\arraybackslash}p{0.16\textwidth}
    >{\raggedright\arraybackslash}p{0.10\textwidth}
    >{\raggedright\arraybackslash}p{0.10\textwidth}
    >{\raggedright\arraybackslash}p{0.15\textwidth}
    >{\raggedright\arraybackslash}p{0.40\textwidth}
@{}}
\toprule
\textbf{Perturbation}
& \textbf{TRAILS}
& \shortstack[l]{\textbf{Context}\\\textbf{relevance}}
& \textbf{Outcome sensitivity}
& \textbf{Interpretation} \\
\midrule
\multicolumn{5}{@{}l}{\textbf{Repeated Prisoner's Dilemma}} \\
\midrule
Persona format
& TRAILS-R & Irrelevant
& Up to 76 pp cooperation shift
& Apparent cooperation equilibria are implementation-sensitive. \\
\addlinespace[3pt]
Game framing
& TRAILS-R & Relevant
& Cooperation and payoff change
& Claims should specify the instruction framing. \\
\addlinespace[3pt]
Memory representation
& TRAILS-R & Irrelevant
& Smaller effects
& Assess preservation against a stated tolerance. \\
\midrule
\multicolumn{5}{@{}l}{\textbf{Echo-chamber simulation}} \\
\midrule
Network homophily
& TRAILS-D & Relevant
& Within-group ratio crosses $0.5$
& Echo-chamber presence is scope-limited by network structure. \\
\addlinespace[3pt]
Hub assignment
& TRAILS-D & Relevant
& Strength changes; presence persists
& Hub composition qualifies echo-chamber strength, not presence. \\
\addlinespace[3pt]
Activation probability
& TRAILS-D & Irrelevant
& No significant changes
& No detected sensitivity in the tested range; not proof of robustness. \\
\addlinespace[3pt]
Memory window
& TRAILS-D & Irrelevant
& No significant changes
& No detected sensitivity in the tested range; not proof of robustness. \\
\addlinespace[3pt]
Recommendation feed size
& TRAILS-D & Irrelevant
& Both metrics increase
& Echo-chamber strength depends on the exposure configuration. \\
\bottomrule
\end{tabular}
\end{table*}

Table~\ref{tab:case-study-perturbations} summarizes GPT-5.2 results under the tested configurations; cross-model results appear in Appendix~\ref{appendix:extended-results}.

\paragraph{Repeated Prisoner's Dilemma.} A claim of stable cooperation is \textit{Implementation-sensitive}: changing persona format alone shifts two-agent cooperation by up to 76 percentage points. Game-instruction framing also changes cooperation and payoff, so the claim should specify its framing conditions. Smaller memory-representation effects require a stated tolerance before claim preservation can be assessed.

\paragraph{Echo-chamber simulation.} Echo-chamber presence is \textit{Scope-limited} by homophily: the weighted same-group edge ratio crosses $0.5$ between the lowest and higher conditions. Hub assignment changes strength, but all conditions remain above $0.5$. Larger feeds strengthen both metrics; activation probability and memory window show no significant effects. Results concern fixed-stance interaction, not opinion polarization; null tests do not establish robustness.



\end{document}